\documentclass{monashthesis}

\usepackage{subcaption}
\usepackage{lscape}
\author{Dewi Lestari Amaliah}
\title{Visualising Multilevel Regression and Poststratification: Alternatives to the Current Practice}
\studentid{31251587}
\def\degreetitle{Business Analytics Creative Activity}
\hypersetup{
     pdfauthor={Dewi Lestari Amaliah},
     pdftitle={Visualising Multilevel Regression and Poststratification: Alternatives to the Current Practice},
     pdfproducer={Bookdown with LaTeX}
}

\bibliography{thesisrefs}

\begin{document}

\pagenumbering{roman}

\titlepage

{\setstretch{1.2}\sf\tighttoc\doublespacing}

\clearpage\pagenumbering{arabic}\setcounter{page}{0}

\hypertarget{abstract}{%
\chapter*{Abstract}\label{abstract}}
\addcontentsline{toc}{chapter}{Abstract}

Surveys provide important evidence for policy making, decision making and understanding society. However, conducting the large surveys required to provide subpopulation level estimates is expensive and time-consuming. Multilevel Regression and Poststratification (MRP) is a promising method to provide reliable estimates for subpopulations from surveys without the amount of data needed for reliable direct estimates. Graphical displays have been widely used to communicate and diagnose MRP estimates. However, there have been few studies on how visualisation should be performed in this field. Accordingly, this study examines the current practice of MRP visualisation using a systematic literature review. This study also applies MRP to estimate the Trump vote share in the U.S. 2016 presidential election using the Cooperative Congressional Election Study (CCES) data to illustrate the implication of current visualisation practices and explore alternatives for improvement. We find that uncertainty is not often displayed in the current practice, despite its importance for survey inference. The choropleth map is the most frequently used to display MRP estimates even though it only shows point estimates and could hinder the information conveyed. Using various graphical representations, we show that visualisation with uncertainty can illustrate the effect of different model specifications on the estimation result. In addition, this study also proposes a visualisation strategy to also take the bias-variance trade-off into account when evaluating MRP models.\\
\newpage

\hypertarget{acknowledgements}{%
\chapter*{Acknowledgements}\label{acknowledgements}}
\addcontentsline{toc}{chapter}{Acknowledgements}

I would like to express my deepest gratitude to my supervisors, Lauren Kennedy and Shiro Kuriwaki. They have given me so much time, knowledge, wisdom, and patience since I started until I finished this project. Their continuous feedback, guidance, encouragement, and advice were like a light in the dark, especially when this project became increasingly challenging to complete.

I would like to extend my gratitude to Dan Simpson, the chief examiner of Master of Business Analytics and Creative Activity, for all the guidance in completing this unit.

I also would like to thank Australia Awards Scholarship for giving me the scholarship to study at the Monash University. Without it, studying in Australia would only remain as one of my childhood dreams.

Last but not least, I dedicate this work to my late father, who always gave me unconditional love and support in pursuing my dreams.

\hypertarget{r-packages}{%
\section*{R packages}\label{r-packages}}
\addcontentsline{toc}{section}{R packages}

Several R \autocite{R} packages are utilized to produce this report: \texttt{mrpkit} \autocite{mrpkit}; \texttt{ccesMRPprep} \autocite{ccesmrpprep}; \texttt{brms} \autocite{brms}; \texttt{cmdstanr} \autocite{cmdstanr}; \texttt{ddi} \autocite{ddi};\texttt{survey} \autocite{survey}; \texttt{tidyverse} \autocite{tidyverse}; \texttt{forcats} \autocite{forcats}; \texttt{Metrics} \autocite{Metrics}; \texttt{data.table} \autocite{datatable}; \texttt{kableExtra} \autocite{kable}; \texttt{janitor} \autocite{janitor}; \texttt{scales} \autocite{scales}; \texttt{ggplot2} \autocite{ggplot2}; \texttt{patchwork} \autocite{pw}; \texttt{flipPlots} \autocite{flipflops}; \texttt{igraph} \autocite{igraph}; \texttt{urbnmapr} \autocite{urban}; \texttt{ggstance} \autocite{stance}; \texttt{ggpmisc} \autocite{ggpmisc}; \texttt{wacolors} \autocite{wacolors}; \texttt{rmarkdown} \autocite{rmd}; \texttt{knitr} \autocite{knitr}; \texttt{MonashEBSTemplates} \autocite{monashtemp}.

\hypertarget{ch:intro}{%
\chapter{Introduction}\label{ch:intro}}

Accurate population and subpopulation estimates are essential to draw insight from the data, especially when policies or decisions are made given the specific context of smaller regions. However, conducting a large survey to provide statistics at a subpopulation level is expensive, time-consuming, and often needs to account for unrepresentative samples. Multilevel regression and poststratification, henceforth referred to as MRP, is a model-based approach used to estimate subpopulations. In short, MRP incorporates a multilevel regression technique to predict the outcome of interest using survey data. This prediction is then poststratified using the population size from a larger survey or census to get the population estimates.

MRP is widely applied to create small area estimates in the absence of a subnational surveys \autocite{hanretty} particularly small geographic areas, such as state or county estimates. MRP also allows the demographic-wise estimation, such as gender, age group, and education. Additionally, MRP is also often applied to adjust the estimation from a non-representative survey as the result of difficulties in recruiting representative survey respondents \autocite{GelmanAndrew2007SwSW}

The standard method to communicate and validate the MRP estimates, such as their accuracy, is by using graphics. Indeed, statistical graphics are regarded a powerful tool to communicate quantitative information and analyse data \autocite{ClevelandWilliamS,1983Gmfd}. \textcite{WickhamHadley2015VsmR} state that statistical visualisation, particularly model visualisation, is imperative as it helps us to understand the model better, for example, how the model changes as its parameters change or how the parameters change as the data changes. They also mention that model visualisation is important to show the model's goodness of fit and whether it is good for some regions only and worse in other regions, or whether it is uniformly good.

While visualisation is common to communicate and diagnose MRP models, there are only a few discussions and studies on how it should be performed. \textcite{mekelagelman} and \textcite{saundra} work on similar areas focussing on the use of graphics in political science. However, \textcite{mekelagelman} only focus on a graphical method for discovery and communication purposes of polling results. Besides, the MRP visualisations that they display as examples are isolated on Gelman's previous papers only. Meanwhile, the latter study by \textcite{saundra} only focuses on how the graphics in public opinion research should be displayed. Therefore, this study tries to fill the gap by discussing the current practices of MRP visualisations generally, not only in public opinion and polling estimates applications. It also aims to explore the possible alternative improvements to current practice.

Explicitly, the objectives of this study are: discuss the current practice of visualisation of MRP models; understand the implication of existing visualisation choices with real-world data; and explore possible improvements of the current practice of MRP visualisation.

The first objectives will be reached by doing a systematic literature review on peer-reviewed articles that applied MRP, while the second and the third goals will be demonstrated through a case study on the 2016 U.S. presidential election using the Cooperative Congressional Election Study (CCES) and the American Community Survey (ACS) data.

\hypertarget{overview}{%
\section{MRP Overview}\label{overview}}

MRP is essentially conducted with two stages - a multilevel/hierarchical regression modeling stage and poststratification stage. The idea is to combine model-based estimation commonly used in small area estimation with poststratification, which is considered the general framework as a weighting scheme in survey analysis \autocite{Gelman97poststratificationinto}. \textcite{Gelman97poststratificationinto} argue that using multilevel regression estimates for poststratification allows the estimation for many more categories to gain more detailed population information.

Formally, let \(K\) be the number of categorical variables in the population and the \(k_{th}\) variable have \(J_k\) categories/levels, the population can be then expressed as \(J = \prod_{k=1}^K J_k\) cells. For every cell, there is a known population size \(N_j\). If the variable in the population is not in categorical form, then it should be converted into a categorical variable first. Next, suppose that the outcome of interest is a binary variable. The MRP procedure is summarised in two stages as follows \autocite{GaoYuxiang2021IMRa}:

\begin{enumerate}
\def\labelenumi{\arabic{enumi}.}
\tightlist
\item
  \textbf{Multilevel regression stage}. Multilevel regression is fitted to get estimated population averages \(\theta_j\) for every cell \(j \in \{1, ...., J\}\). The multilevel logistic regression has a set of random effects \(\alpha^k_{m[j]}\) for each categorical covariate \(k\). These random effects have the effect of pooling each \(\alpha_j\) partially towards overall grand mean. Suppose that \(n\) is the number of individual observations in the survey data, the form of multilevel regression could be written as follows:
\end{enumerate}

\begin{equation}
\begin{split}
& Pr(y_i = 1) = logit^{-1}\left(X_i\beta + \sum^K_{k=1}\alpha^k_m[i]\right), for\ i=1, ..., n,\\
& \alpha^k_m \sim N(0, \sigma^2_k), for\ m = 1, ..., M_k
\end{split}
\label{eq:mrp-stage1}
\end{equation}

\begin{enumerate}
\def\labelenumi{\arabic{enumi}.}
\setcounter{enumi}{1}
\tightlist
\item
  \textbf{Poststratification stage}. The probabilities of the outcome in each cell from the previous stage, \(\theta_j\), is then poststratified using the known population size \(N_j\) of each cell \(j\) to get the estimates at the subpopulation level. This stage corrects the nonresponse in the population by utilizing the known size of every cell \(j\) relative to the total population size \(N = \sum_{j=1}^J N_j\). In other words, the estimates is a weighted average of \(\theta_j\) with \(N_j\) as the weight. Suppose that \(S\) is the subpopulation which is the combination of categories in the poststratification matrix, the MRP estimates could be expressed as:
\end{enumerate}

\begin{equation}
\begin{split}
\theta_S = \frac{\sum_{j \in S}N_j\theta_j}{\sum_{j\in S}N_j}
\end{split}
\label{eq:mrp-stage2}
\end{equation}

\hypertarget{report-structure}{%
\section{Report Structure}\label{report-structure}}

The first chapter of this report is introduction in which the motivation and objectives of this study are articulated. Chapter \ref{ch:syslitrev} is a systematic literature review. This chapter discuss the review of current practice in MRP visualisations in various studies. Next, Chapter \ref{ch:case-stud} is a case study of MRP visualisations. This chapter aims to demonstrate the MRP application in the case of U.S. presidential voting result estimation. This chapter also demonstrates how the current practice of MRP visualisation could be improved. The final chapter, Chapter \ref{ch:conclusion}, summarises the findings and concludes the contribution of this study and possible future works.

\hypertarget{ch:syslitrev}{%
\chapter{Systematic Literature Review}\label{ch:syslitrev}}

This study is performed using a systematic review method. This method collects empirical evidence explicitly and systematically using pre-specified eligibility criteria to answer a specific research question \autocite{cochrane}. Systematic literature reviews also enable the process of finding the gap in a field of science, such as understanding what has been done and what needs to be done \autocite{LinnenlueckeMartinaK2020Cslr}. Hence, in this case, systematic literature review could assist us to understand the common practice in MRP visualisations so that we can explore how to improve.

According to \textcite{brown_uni}, the key criteria of the systematic literature review are: \emph{``a clearly defined question with inclusion \& exclusion criteria; rigorous \& systematic search of the literature; critical appraisal of included studies; data extraction and management; analysis \& interpretation of results; and report for publication.''} Hence, to conform with these criteria, this study incorporates the Preferred Reporting Items for Systematic Reviews and Meta-Analysis (PRISMA)'s checklist and flow diagram. The following subsections discuss the steps conducted following these criteria.

\hypertarget{literature-identification}{%
\section{Literature Identification}\label{literature-identification}}

MRP is applied in various scientific fields, ranging from social and political science to public health. Therefore, to identify relevant literature, this study refers to research databases instead of field-specific journals. Those databases are JSTOR, EBSCO, and PubMed. The first two databases are chosen due to their broad range of field coverage, while the latter is chosen since MRP is sometimes also applied in the health and medical fields. These databases were also chosen to represent the heterogeneity of the field, which is one of the important factors in a systematic literature review \autocite{SchweizerMarinL2017Apgt}.

From these databases we identify relevant articles using the combination of several search terms. Generally the search terms include the term ``multilevel regression'', ``post-stratification'', ``poststratification'', and ``multilevel model''. Our target literature is articles that are written in English. We exclude all of the publications before 1997 since this was the first proposal date for MRP. Initially we included only the title/abstract when searching these databases. However, using this method limits the set of potential articles to only include those with the search term in the abstract/title. To rectify this, we also include a search with ``all field'' in the search criteria. Note that for EBSCO, we directly apply the search for all fields. The detailed literature identification is shown in Table \ref{tab:search-term}.

The total number of articles from this search criteria are 327. Next, we utilize the literature manager, EndNote X9, to manage these articles and to find duplicate articles. After removing those duplicate articles, we have 212 articles to be screened in the next stage.

\begin{landscape}\begin{table}

\caption{\label{tab:search-term}Detail of literature identification}
\centering
\resizebox{\linewidth}{!}{
\begin{tabular}[t]{l>{\raggedright\arraybackslash}p{15em}lllr}
\toprule
Database & Search Terms & Search Field & Inclusion & Exclusion & Number Returned\\
\midrule
JSTOR & (multilevel regression and poststratification) OR (“post-stratification”) & Abstract & Article, content I can access, English & anything before 1997 & 44\\
JSTOR & (("multilevel regression" AND ("post-stratification" OR Poststratification)) OR ("multilevel model" AND ("post-stratification" OR Poststratification))) & All field & Article, English & anything before 1997 & 142\\
EBSCO & "multilevel regression with post-stratification" OR "multilevel regression with poststratification" OR "multilevel regression and Poststratification" OR "multilevel regression and Post-stratification" & All field & Academic (Peer-Reviewed) Journals, English & anything before 1997 & 42\\
EBSCO & (multilevel regression AND post-stratification) OR (multilevel model AND post-stratification) OR (multilevel regression AND poststratification ) OR (multilevel model AND poststratification) & All field & Academic (Peer-Reviewed) Journals, English & anything before 1997 & 45\\
PubMed & "multilevel regression with post-stratification" OR "multilevel regression with poststratification" OR "multilevel regression and Poststratification" OR "multilevel regression and Post-stratification" & Title/Abstract & Article, English & anything before 1997 & 26\\
\addlinespace
PubMed & (multilevel regression AND post-stratification) OR (multilevel model AND post-stratification) OR (multilevel regression AND poststratification) OR (multilevel model AND poststratification) & All field & Article, English & anything before 1997 & 28\\
\bottomrule
\end{tabular}}
\end{table}
\end{landscape}

\hypertarget{screening-and-eligibility-criteria}{%
\section{Screening and Eligibility Criteria}\label{screening-and-eligibility-criteria}}

We screen all of the articles based on predetermined criteria. We find that 3 articles are apparently not research papers. This results in 209 abstracts to be screened. To screen efficiently, we use two stages. The first stage is a review of abstracts, the second a full manuscript review.

\hypertarget{stage-1-review-of-abstracts}{%
\subsection{Stage 1: Review of abstracts}\label{stage-1-review-of-abstracts}}

In the first stage the author and and my supervisor (Kennedy) independently review all article abstracts with the following eligibility criteria:

\begin{enumerate}
\def\labelenumi{\arabic{enumi}.}
\tightlist
\item
  The abstract should mention analysis of data or creation of simulation data.
\item
  The abstract should mention the use of MRP or multilevel models to make population estimates or the use of other regression models (BART, spatial, stacking, trees) to make population estimates.
\end{enumerate}

During the screening, we agreed agreed that 61 articles meet the eligibility criteria listed above, while 104 articles do not meet the criteria. The two reviewers disagreed on 44 articles. Accordingly, we skim the full manuscript to decide whether the paper could be included in the next stage or not. As the result, an additional 22 more articles are moved to stage 2, making a total of 83.

\hypertarget{stage-2-full-manuscript-review}{%
\subsection{Stage 2: Full manuscript review}\label{stage-2-full-manuscript-review}}

DA reviews the full manuscript on 83 articles based on a second set of criteria. The aim of this stage is to get the list of the final articles that would be included in the study. We set the criteria of inclusion as follow:

\begin{enumerate}
\def\labelenumi{\arabic{enumi}.}
\tightlist
\item
  It should apply MRP as its method.
\item
  It should contain at least one plot relate to MRP findings.
\end{enumerate}

\begin{figure}
\includegraphics[width=0.9\linewidth]{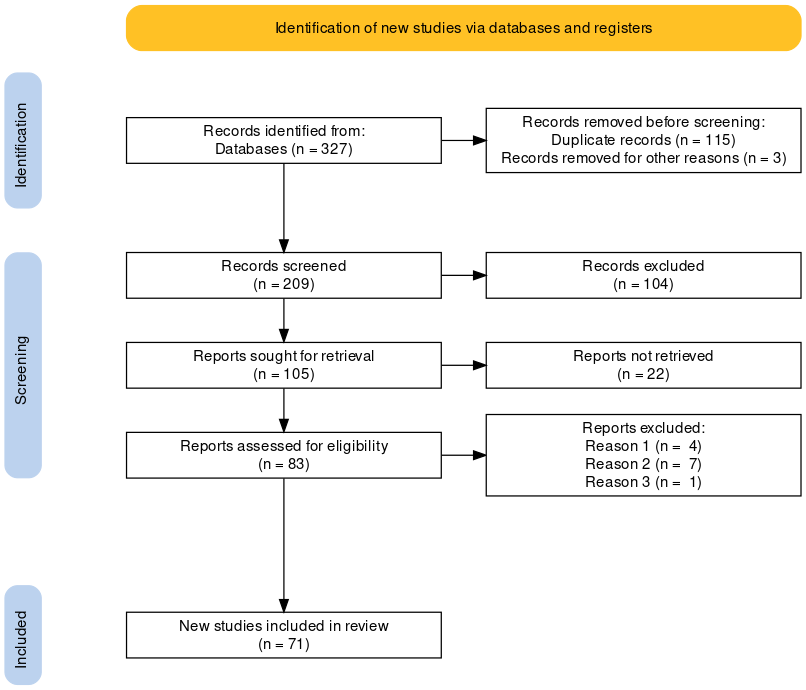} \caption{PRISMA flow chart of this systematic literature review.}\label{fig:prisma-flowchart}
\end{figure}

During this stage, we exclude 4 articles as they do not meet the first criteria. Further, 7 articles are excluded as they do not meet the second criteria. Also, an article is not included because it is a duplicate that was not detected automatically by Endnote X9. Finally, we have 71 articles to be reviewed in the next stage. Figure \ref{fig:prisma-flowchart} displays the PRISMA flow chart of this study. This figure is generated using \texttt{PRISMA2020} \autocite{prisma2020}.

\hypertarget{data-extraction-and-analysis}{%
\section{Data Extraction and Analysis}\label{data-extraction-and-analysis}}

We focus the data extraction on the MRP-related plot. We manually create a metadata for each plot (included in the supplementary material). We will use this metadata to analyse the current reporting practices with MRP. This metadata will also ensure the reproducibility of the analysis and to maintain the transparency of the systematic literature review process.

We code the plots according to their type, i.e., communication (coded to 0) and diagnostic plot (coded to 1). For diagnostic plots, we examine whether the plots compare MRP with other estimates, which are:

\begin{enumerate}
\def\labelenumi{\arabic{enumi}.}
\tightlist
\item
  Raw (direct estimates or direct disaggregation);
\item
  Ground truth;
\item
  Weighted estimates;
\item
  Estimates from other MRP models, for example, a paper build several MRP models from various simulation scenarios or using different covariates;
\item
  Estimates from another study/survey;
\item
  Estimates from another method, for example comparing MRP with Bayesian Additive Tress with Post-Stratification(BARP).
\end{enumerate}

Plots that show a comparison of MRP with the above list would be coded to 1, otherwise coded to 0.
Diagnostic plots also categorised based on how they compare the performance of MRP. The five observed criteria are:

\begin{enumerate}
\def\labelenumi{\arabic{enumi}.}
\tightlist
\item
  Bias;
\item
  Mean Absolute Error (MAE);
\item
  Mean Square Error (MSE)/ Relative Mean Square Error (RMSE);
\item
  Standard Error (SE);
\item
  Correlation.
\end{enumerate}

Each plot is assessed based on the use of the performance metric. For each metric is scored based on whether it is used (coded 1) or not (coded 0).

We also review other features of the plot using the grammar in \texttt{ggplot2} \autocite{ggplot2} as a framework. The common grammar used in practice allows us to understand to what extend MRP models are effectively visualised. It is worth noting that there is no specific convention or well-documented recommendation on how data should be visualised as building a graph more often involves choice or preference \autocite{MIDWAY2020100141}. For example, there is no specific convention on which variable should be put on the x and y-axis in a scatter plot, even though it has been common knowledge to put the response variable on the y-axis and the explanatory variable on the x-axis. Hence, grammar assists us in evaluating well-formed graphics \autocite{layered-grammar}. In addition, \textcite{vanderplas} mention that classifying and comparing graphs according to their grammar is more robust and more elegant.

Accordingly, we examine the facet, geom, axis, color, and shape. For reproducibility, the metadata also contains the article's author/s, publication year, title, and corresponding figure number as it appeared in the article. After the extraction, we analyze the data using graphical visualization with \texttt{ggplot2} \autocite{ggplot2}. The result will be discussed in the following subsection.

\hypertarget{com-prac}{%
\section{Common practices in MRP visualisations}\label{com-prac}}

In this study, graphics are classified into two types, i.e., communication and diagnostic plots. A plot is classified as a communication plot if the plot's goal is solely to convey the MRP result. A diagnostic plot is used to understand the MRP estimate, and typically displays the MRP estimation by showing the performance metrics or compares it with other estimation methods. From 71 articles, we extract the data of 243 plots. 47.33 \% of these plots are diagnostics plots, while the remaining are communication plots.

\hypertarget{performance-metrics-used-in-mrp}{%
\subsection{Performance metrics used in MRP}\label{performance-metrics-used-in-mrp}}

According to \textcite{BotchkarevAlexei2019ANTD}, performance metrics is \emph{``a logical and mathematical construct designed to measure how close are the actual results from what has been expected or predicted''} RMSE and MAE are among the most common methods used in many studies \autocite{BotchkarevAlexei2019ANTD}. However, \textcite{WillmottCJ2005Aotm} states that RMSE should not be reported in any studies since it could be multi-interpreted because it does not describe average error alone and MAE is more appropriate metric. This argument is denied by \textcite{ChaiT2014Rmse} who argue that RMSE is not ambiguous and better than MAE if the distribution of model's error is normal. Accordingly, there is no single metric that fits all methods \autocite{ChaiT2014Rmse}.

\begin{figure}
\centering
\includegraphics{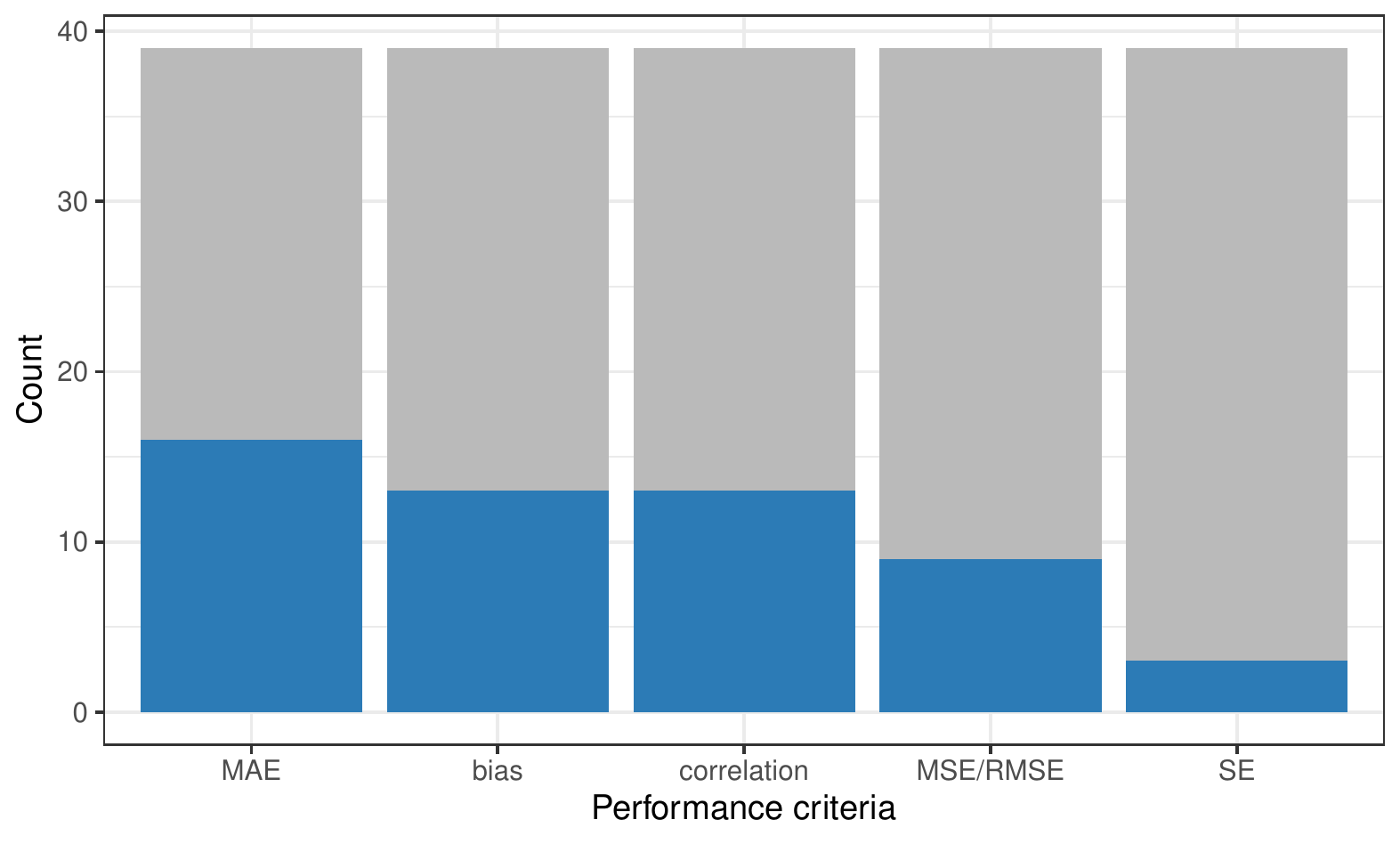}
\caption{\label{fig:perform-plot}We observed five performance metrics used: Mean Absolute Error (MAE), bias, correlation, Mean Square Error/Root Mean Square Error (MSE/RMSE), and Standard Error (SE). Each bar represents the number of plot that show performance metrics, particularly, the grey shade represents the number of plot that show MRP performance but did not use the corresponding metrics. It is possible that a plot shows more than one metrics, so that the blue bars do not count to the sum. We learn that MAE is metrics that is mostly shown in plots we reviewed.}
\end{figure}

In this study, we find that there are 39 plots out of 115 diagnostic plots (about 34\%) that display performance measures. As seen in Figure \ref{fig:perform-plot}, we find that MAE is the most widely used performance metric in MRP visualisations. Bias, which is interpreted similarly to MAE, is also widely used. Meanwhile, the squared error measures, which are MSE/RMSE and standard error, are only used in a few plots. It is interesting that correlation, which is not a common metric for performance, is more widely used than square error metrics.

Most of these metrics only refer to point estimates, i.e., the distance between the predicted value and the actual values. Also, these metrics mainly measure bias. However, MRP is a model in which bias-variance is applied. Therefore, other measures are also needed that reflect the degree of uncertainty and variations in the predicted value. Measures such as length of confidence or credibility interval can be used, in which the narrower the value, the more precise the estimates.

\hypertarget{common-comparisons-with-mrp}{%
\subsection{Common comparisons with MRP}\label{common-comparisons-with-mrp}}

The goal of MRP is to make a population estimate. The method aims to adjust an unrepresentative survey to obtain accurate population and sub-population estimates. Where possible MRP is usually compared with a true value. This is generally only possible in political science applications where an election provides this true estimate. To understand how MRP improves estimates from an unrepresentative survey when compared with no adjustment, MRP estimates are usually compared with direct estimates (raw). Similarly, to understand the improvement of estimates when compared with more traditional methods, MRP is often compared with weighted estimates.

\begin{figure}
\centering
\includegraphics{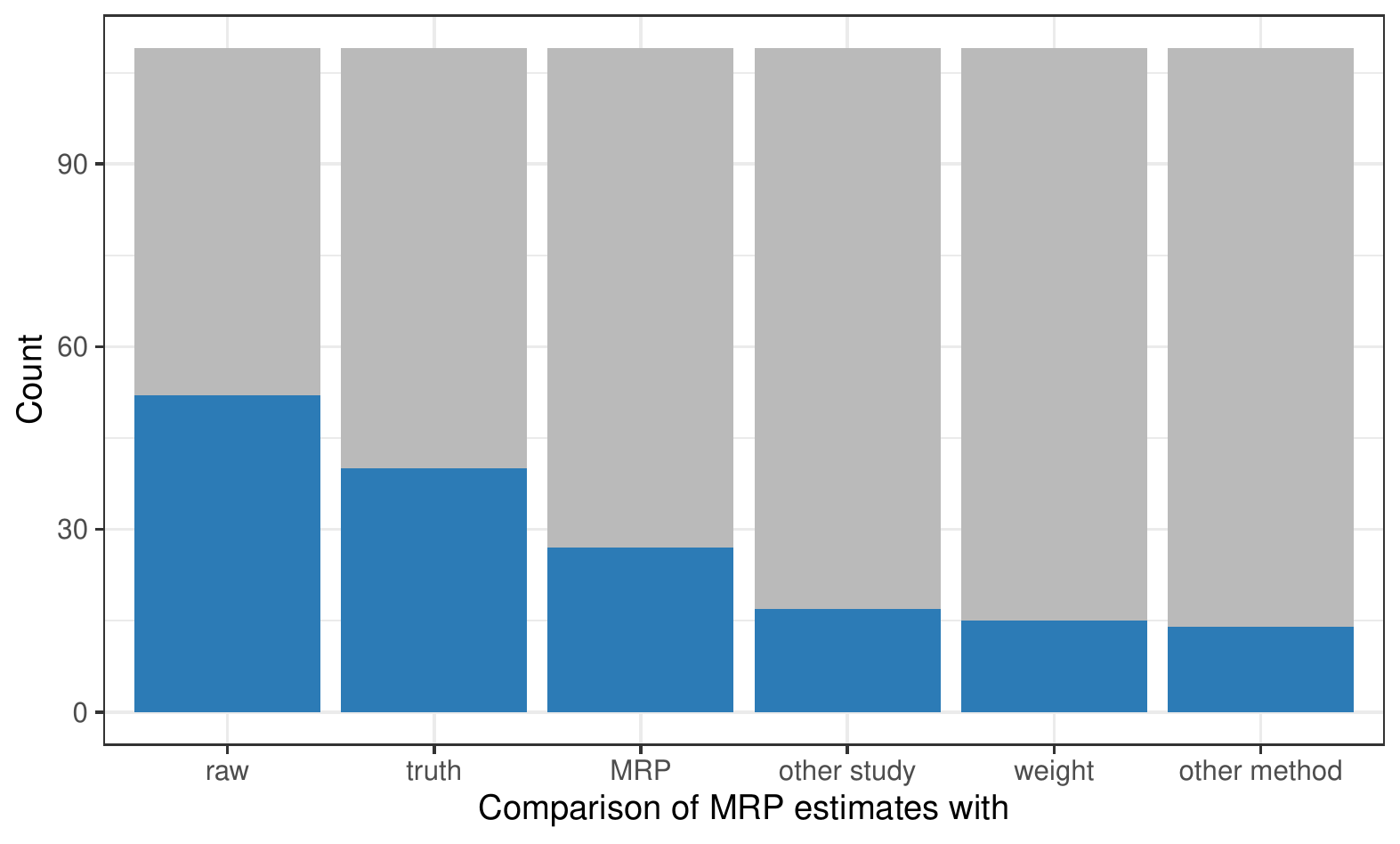}
\caption{\label{fig:compare-plot}Estimates that are compared with MRP. The bars represent the number of plots that display comparison of MRP with other estimates. Particularly, the blue shade represents the number of plots that compare MRP estimates with the estimates shown in each bar, while the grey shade represents the number of plots that also show comparison of MRP but did not compare to this particular estimate. Note that the blue bars do not sum to the count because some plots compared to multiple alternative estimates. It is shown that MRP estimates are mostly compared with raw estimates.}
\end{figure}

This study finds that from 115 diagnostic plots, 109 (about 95\%) compare MRP estimates with estimates from other methods. Figure \ref{fig:compare-plot} shows the distribution of alternative estimates. MRP estimates are mostly compared to direct estimates and the ground truth. Some studies also compare estimates from several MRP models (usually with different model specifications). There are not many plots showing the comparison between MRP estimates and weighted estimates.

\hypertarget{common-grammar-in-mrp-visualisations}{%
\subsection{Common grammar in MRP visualisations}\label{common-grammar-in-mrp-visualisations}}

\textbf{Plot type}

\begin{figure}
\centering
\includegraphics{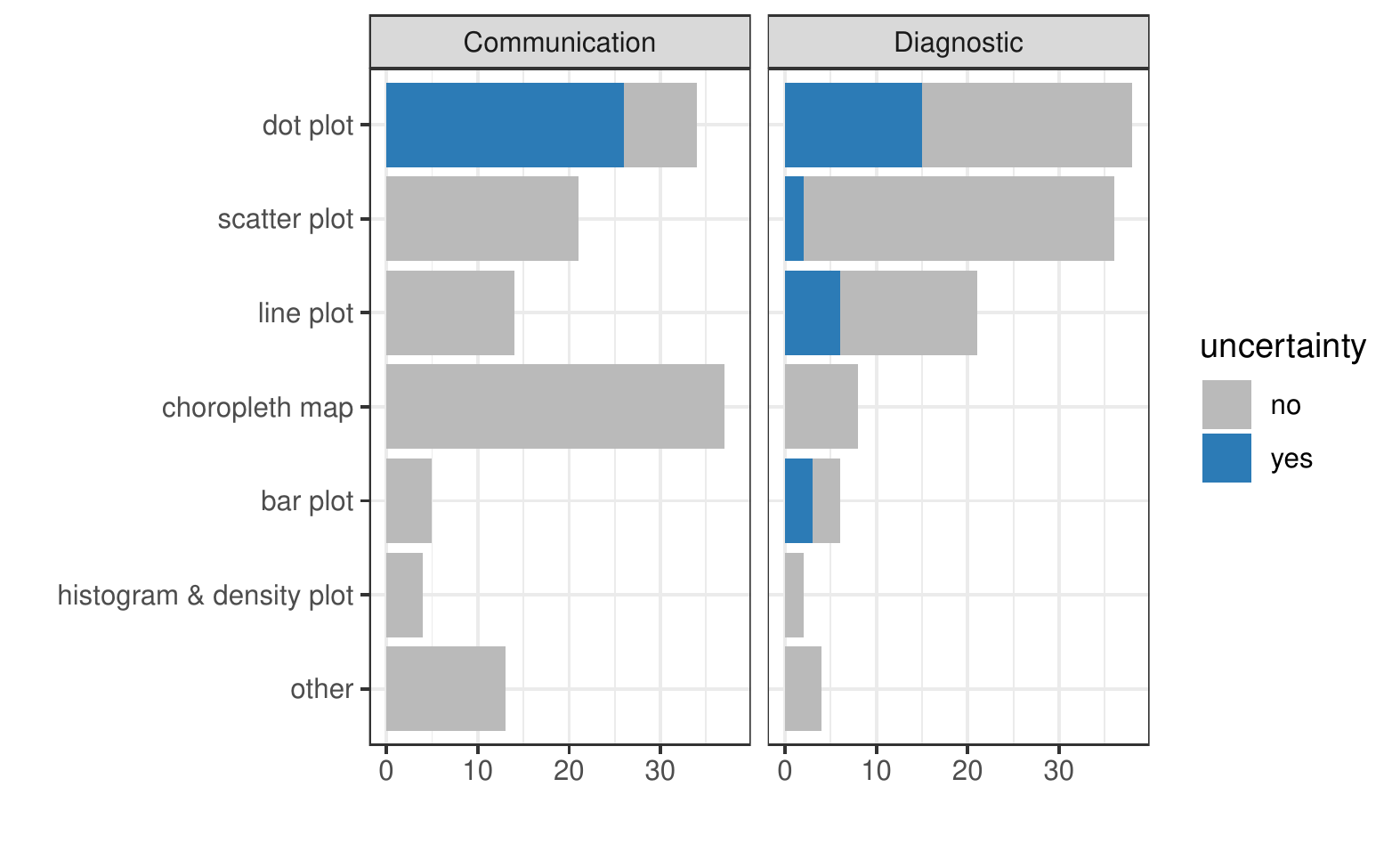}
\caption{\label{fig:common-plots}Common plot types used in MRP visualisations. The blue shade display the number of plots that showed uncertainty, while the grey shade display the number of plots that did not show uncertainty. Both communication and diagnostics plots rarely displayed uncertainty.}
\end{figure}

Plot type, referred to as \texttt{geom} in the grammar of graphics, represents the shape and features displayed in the graph. Figure \ref{fig:common-plots} suggests that communication and diagnostic plots have a different pattern in which plot types are used (See Appendix \ref{terms} for description/definition of each plot type). Communicating MRP estimates are mostly done using a choropleth map as MRP is often used for small area estimation. For diagnostic purposes, dot plots are mostly used to compare more than two estimation methods or to show some performance metrics.

Notice that Figure \ref{fig:common-plots} also displays the use of uncertainty in MRP model visualisations. According to \textcite{MIDWAY2020100141}, displaying uncertainty in the statistical graphs is essential as the absence of this measure would produce a misleading interpretation and hinder some statistical messages. However, he further states that uncertainty is often neglected in data visualisation. This is what we find in this study - uncertainty is not often seen in the plots. This is possibly because many of the application areas are more familiar with official statistics. In official statistics uncertainty is often unreported because results that are not sufficiently precise are not reported.

\textbf{Values put in x and y-axis}

\begin{figure}
\centering
\includegraphics{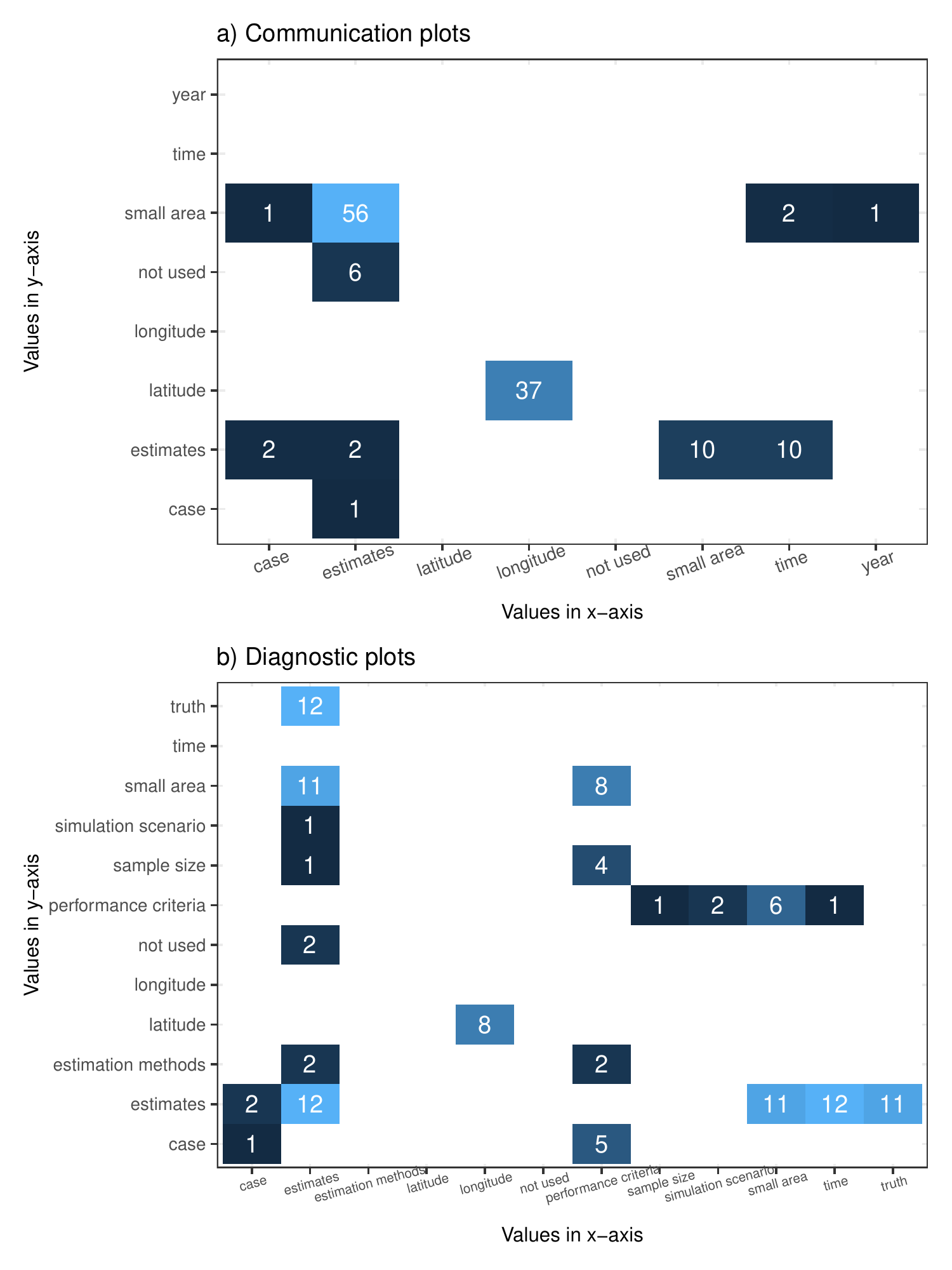}
\caption{\label{fig:common-axis}Common values put in plots' axis. If the values represented in the x and y-axis are longitude and lattitude, it means that the plot is a map. `not used' means that the plot is one dimension. It conveys that axis in diagnostic plots more varied compared to communication plot.}
\end{figure}

The main component of a data visualisation is the axis. x and y-axis represents what value/data are exactly displayed in the graph. In MRP visualisations (Figure \ref{fig:common-axis}), estimates, small area, actual value (truth), and time are among the values that are displayed in the plot. We can also see that the constructs represented by the x and y-axis are more varied in diagnostic plots. It is worth noting that there are no strict rules on values to put in x and y-axis. However, it is a common that the the fixed value is represented by the x-axis, while the random variable is represented in the y-axis. We do not see this in our results as we find estimates and truth are plotted on the x and y axes interchangeably. Another common rule of thumb is that time is almost always represented on the x-axis, which is supported by the findings of our study.

\textbf{Facet}

\begin{figure}
\centering
\includegraphics{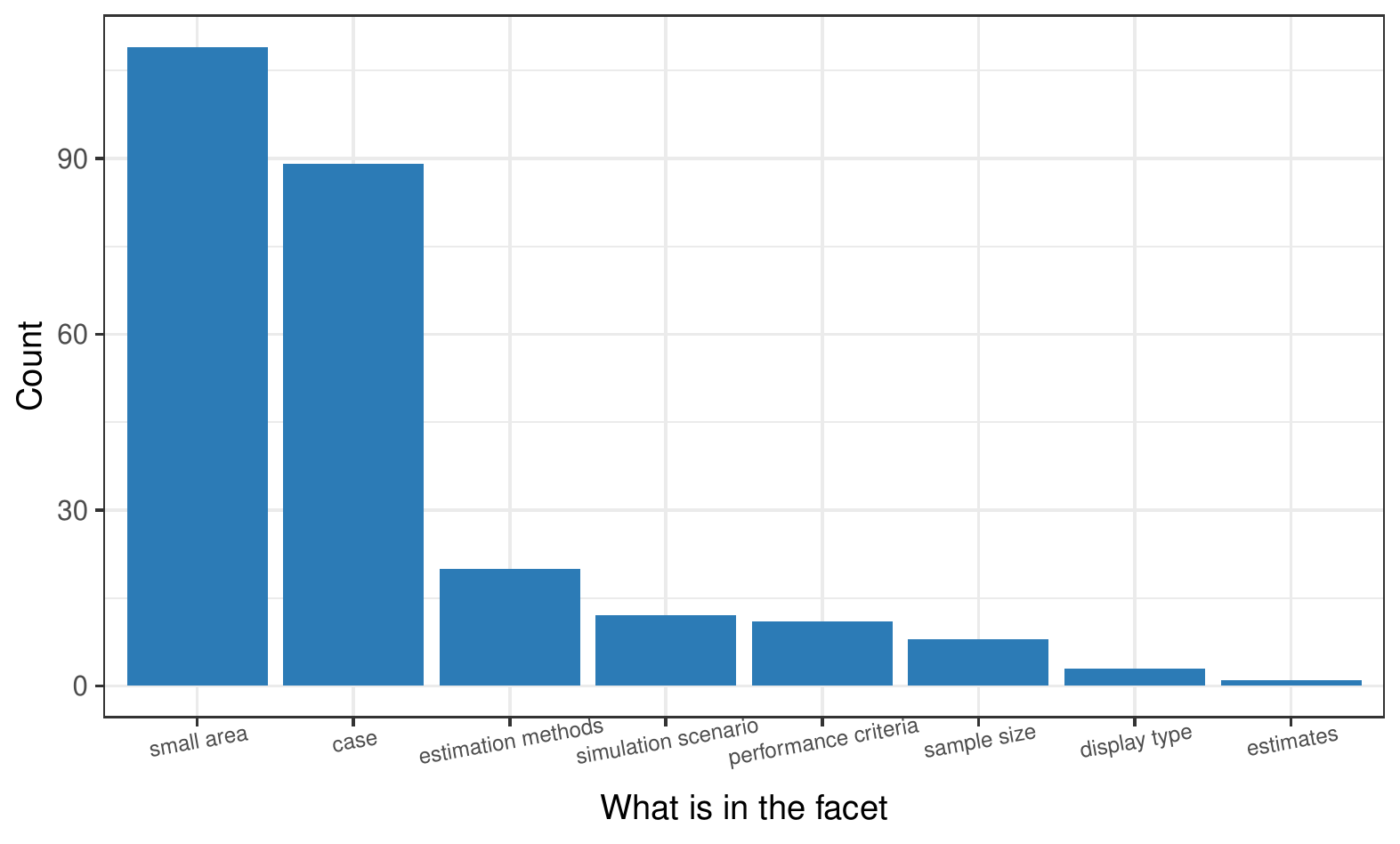}
\caption{\label{fig:facet-plots}The facetting variable in MRP visualisations. Most of plots in the articles reviewed are faceted by small area and case (outcomes measured)}
\end{figure}

Paneling or faceting is considered as to one of the effective visualisation techniques to compare the same variables by its grouping factor \autocite{MIDWAY2020100141}. We find in our results that faceting is a common practice in MRP visualisations. Figure \ref{fig:facet-plots} shows that faceting the plots by small area that is being estimated is the most common, followed by case. Small area refers to the levels of the predictors in the MRP model, for example, state, county, and religion. In several plots, small area could be referred to another variable that is associated with the MRP estimates, but is not included in the model, such as the association between health literacy and the opinion on a health-related bill. Health literacy is a variable that is not included in fitting the MRP model, while the latter is the MRP estimates. Further, case is referred to the outcome predicted with MRP.

\textbf{Other features used}

Besides the features explained previously, color and shape are also the components of grammar of graphics. According to a large experimental study on visualisations, color is a memorable feature of a graph \autocite{MIDWAY2020100141}. Further, \textcite{few_2008} states that the aim of color in data visualisations are to highlight particular data, to group items, and to encode quantitative values. In addition, color is sometimes displayed along with shapes to distinguish more features.

\begin{figure}
\centering
\includegraphics{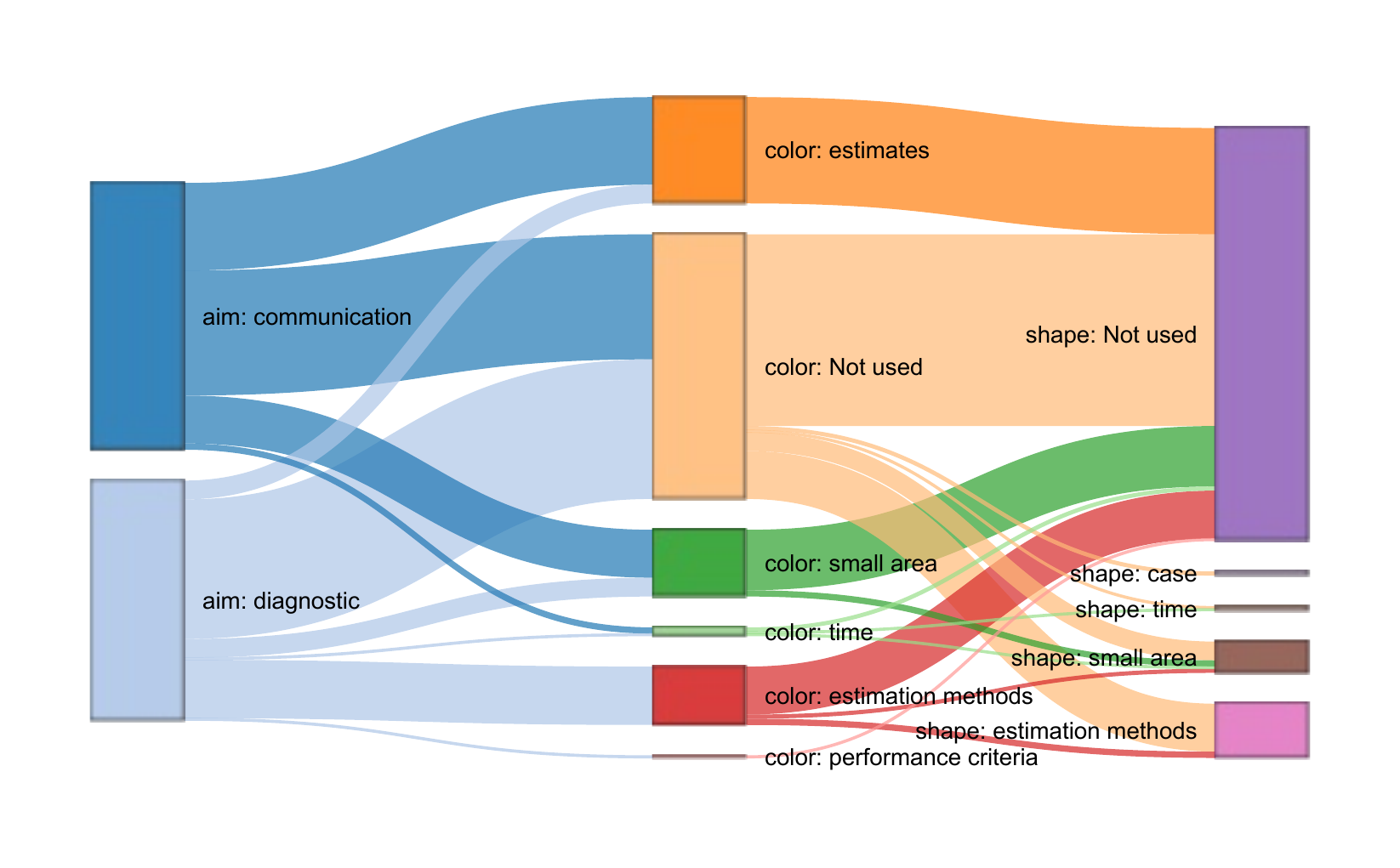}
\caption{\label{fig:sankey-feature}Values that are commonly represented by color and shape in MRP visualisations. Both communication and diagnostic plots rarely use color and shape features to display values.}
\end{figure}

We find, as shown in Figure \ref{fig:sankey-feature}, that both communication and diagnostic plots incorporate color only about half the time. Shape is used less often. When there is only one feature to be displayed, for example, estimation methods, people tend to choose to use color first, rather than shape. This is seen in Figure \ref{fig:sankey-feature} as after incorporating color to distinct estimates, small area, estimation methods, and performance criteria, people tend to not use shape anymore.

\hypertarget{ch:case-stud}{%
\chapter{Case Study: Application of MRP in Presidential Voting Estimation}\label{ch:case-stud}}

The majority of MRP applications are used in the context of estimating public opinion in the social and political sciences, although, in recent developments MRP has also been used in other fields, for example, health and environmental studies. When first introduced by \textcite{Gelman97poststratificationinto}, MRP was applied to generate state estimation of the 1988 U.S. presidential election. Various subsequent studies also made presidential voting the case of interest. We recorded at least seven articles (\textcite{GelmanAndrew2014HBAC}; \textcite{GhitzaYair2013DIwM}; \textcite{KiewietdeJongeChadP2018PSPE}; \textcite{LauderdaleBenjaminE2020Mppf}; \textcite{LeiRayleigh2017T2EA}; \textcite{ParkDavidK2004BMEw}; \textcite{WangWei2015Fewn}) included in the systematic literature review in Chapter 2 that also applied MRP to presidential election estimation. In this chapter, we will also apply MRP to estimate the 2016 U.S. presidential voting outcome, specifically the probability of voting for Donald Trump in this election. This also allows us to compare MRP estimates with the actual value of the Trump votes that are already available. In this case study, we use the Cooperative Congressional Election Study (CCES) 2016 data \autocite{cces_data} as the survey data and the American Community Survey data 2015-2017 \autocite{acs_data} as the population/ poststratification data.

\hypertarget{data}{%
\section{Data}\label{data}}

\hypertarget{cooperative-congressional-election-study-cces-2016}{%
\subsection{Cooperative Congressional Election Study (CCES) 2016}\label{cooperative-congressional-election-study-cces-2016}}

CCES is an annual survey that aims to capture Americans' view on Congress, their voting behavior and experience with regards to political geography, social, and demographic context \autocite{cces_data}. In 2016, the CCES covers 64,600 samples spread over 51 states. Accordingly, \textcite{cces_data} suggest that the data is precise enough to measure the distribution of voters' preference in most states. In addition, beyond it's large sample size, CCES is regarded to be a desirable dataset because it measures vote preference before and after the election so that it is more reliable compared to a single question format \autocite{kuriwaki}.

To fit MRP models, we use several variables from this survey. To obtain the data from the CCES website, we utilize an R package, \texttt{ccesMRPprep} \autocite{ccesmrpprep}. One of the advantages of using this package is that the data has been pre-processed in particular for MRP purposes, in this case, we use the \texttt{ccc\_std\_demographics} function. Also, the variable names are already recoded so it has more interpretable names. The code to get the data is available in the supplementary materials of this report.

Throughout this demonstration, we estimate the proportion of voters who turned out to vote for Trump in the 2016 U.S. Presidential Election. We choose this outcome following other demonstrations (e.g.~\textcite{kuriwaki}; \textcite{MengXiao-Li2018Spap}) precisely because the population quantity is observed after the survey is run. That allows us to validate my estimates against a ground truth. To visualise the implication of different model specification, we also choose other two outcomes, which are vote preference and party identity. In CCES 2016 these variables named candidate voted for (\texttt{CC16\_410a}), the vote preference/intention (\texttt{CC16\_364c}),and party identity (\texttt{pid3} including leaners who are coded as Independents in pid3 but expressed leaning towards a party in \texttt{pid7}). Table \ref{tab:outcome-table} shows the distribution of answers in those three variables. In \texttt{ccesMRPprep}, these variables have been renamed to \texttt{intent\_pres\_16}, \texttt{voted\_pres\_16}, and \texttt{pid3\_leaner}, respectively. It is worth noting that the MRP models we would like to build use binary responses. As we are comparing to the US presidential election, we would like a variable that represents whether the respondents vote for Trump/Republican or not.

\begin{table}
\caption{\label{tab:outcome-table}Percentage of each answer in CCES 2016 (n = 64,000). This question will be the MRP models outcome in this case study. Since the model outcome is binary, these answer will be converted to be yes/no in the context of vote for Trump/Republican.}

\centering
\begin{tabular}[t]{lr}
\toprule
Candidate voted & percentage\\
\midrule
Hilary Clinton & 34.27\\
Donald Trump & 29.03\\
Other / Someone Else & 6.26\\
Did Not Vote & 0.13\\
Not Sure / Don't Recall & 0.35\\
\addlinespace
NA & 29.97\\
\bottomrule
\end{tabular}
\centering
\begin{tabular}[t]{lr}
\toprule
Candidate will be voted & percentage\\
\midrule
Donald Trump (Republican) & 29.76\\
Hillary Clinton (Democrat) & 42.57\\
Gary Johnson (Libertarian) & 4.87\\
Jill Stein (Green) & 2.17\\
Other & 2.91\\
\addlinespace
I Won't Vote in this Election & 5.13\\
I'm Not Sure & 10.12\\
NA & 2.47\\
\bottomrule
\end{tabular}
\centering
\begin{tabular}[t]{lr}
\toprule
Party identity including leaners & percentage\\
\midrule
Democrat (Including Leaners) & 48.20\\
Republican (Including Leaners) & 32.27\\
Independent (Excluding Leaners) & 16.24\\
Not Sure & 3.20\\
NA & 0.08\\
\bottomrule
\end{tabular}
\end{table}

Further, the geography and demographic variables used as covariates in the models are \texttt{state}, \texttt{age}, \texttt{gender}, \texttt{education}, and \texttt{race}. Table \ref{tab:covariate-tables} shows the distribution of categories/levels of \texttt{age}, \texttt{gender}, \texttt{education}, and \texttt{race}. Initially, age recorded as integers but we transformed it into five age groups. Also, \texttt{education} and \texttt{race} have more levels in the original data but are collapsed to have to obtain fewer levels. In particular, we use the standard/default categorisation in the \texttt{ccesMRPprep} package. The proportion of people answered the survey based on the state is displayed in the appendix of this report (\ref{apd-state}).

\begin{table}
\caption{\label{tab:covariate-tables}The response of covariates. Note that this response has been categorised into certain levels that are reflected in these tables.}

\centering
\begin{tabular}[t]{lr}
\toprule
Gender & percentage\\
\midrule
Male & 45.71\\
Female & 54.29\\
\bottomrule
\end{tabular}
\centering
\begin{tabular}[t]{lr}
\toprule
Race & percentage\\
\midrule
White & 69.44\\
Black & 12.00\\
Hispanic & 10.59\\
Asian & 3.53\\
Native American & 0.81\\
\addlinespace
All Other & 3.63\\
\bottomrule
\end{tabular}
\centering
\begin{tabular}[t]{lr}
\toprule
Age & percentage\\
\midrule
18 to 24 years & 8.30\\
25 to 34 years & 19.62\\
35 to 44 years & 15.75\\
45 to 64 years & 38.36\\
65 years and over & 17.98\\
\bottomrule
\end{tabular}
\centering
\begin{tabular}[t]{lr}
\toprule
Education & percentage\\
\midrule
HS or Less & 28.41\\
Some College & 35.38\\
4-Year & 23.04\\
Post-Grad & 13.17\\
\bottomrule
\end{tabular}
\end{table}

\hypertarget{american-community-survey-acs-2015-2017}{%
\subsection{American Community Survey (ACS) 2015-2017}\label{american-community-survey-acs-2015-2017}}

In this study, we use the ACS 2015-2017 data as the poststratification data. The ACS is a large, survey of the American population conducted by the census bureau and covering jobs and occupations, demographic and citizenship, educational attainment, homeownership, and other topics \autocite{acs_data_about}. The ACS uses monthly probabilistic samples to produce the annual estimates. The ACS is desirable data to represent the U.S. population since the coverage rate, a measure on how well does the survey cover population, for the 2015-2017 ACS is 92.4\%, 91.9\%, 91.6\%, respectively \autocite{acs_coverage_rate}. However, it is also worth noting that the population of interest of the ACS (American population) and the population we are interested in (voting population) is different (the ACS measures the general US population, while the CCES wants to study the behavior of the U.S. adult citizens who turned out to vote), and therefore, bias might always be presented.

To construct the desired poststratification matrix, we need the individual data of the ACS instead of the aggregated statistics. To do this, we use the 1-year Public Use Microdata Sample (PUMS), which carries the information/records of individual people on a yearly basis, appropriately deidentified. The 1-year PUMS data reflects approximately one percent of the U.S. population \autocite{pums_metadata}. Therefore, in this study, we use three years periods of the ACS 1-Year PUMS from 2015-2017 instead of 2016 only to get a better and more stable representation of the American population. Every individual in the data has a weight (\texttt{PWGTP}). Since we use three years period, this weight is then divided by 3 to obtain a population total that matches the full population total.

The data is publicly available on the \href{https://www.census.gov/programs-surveys/acs/microdata/access.2015.html}{U.S. Census Bureau website}. We downloaded the data in a .csv format (csv\_pus.zip) year by year (2015-2019) through access on \href{https://www2.census.gov/programs-surveys/acs/data/pums/2015/1-Year/}{FTP site}. After that, we did a data pre-processing to bind the three years of the PUMS data. We only use some variables in this data for the MRP-purposes, i.e., unique identifier of the person (\texttt{SERIALNO}), state (\texttt{ST}), weight (\texttt{PWGTP}), education (\texttt{SCHL}), sex (\texttt{SEX}), race (\texttt{RAC1P}), Hispanic origin (\texttt{HISP}), and age (\texttt{AGEP}). We also did a data munging to recode and collapse some categories in these variables. Note that the \texttt{RAC1P} did not record for Hispanic ethnicity. Hence, we introduce a new category here, Hispanic, identified if the person answers other than ``1'' in the \texttt{HISP} variable. Table \ref{tab:acs-response-freq} shows the categorised response of the variables obtained from the ACS, i.e., age, race and ethnicity, and education (see Appendix \ref{apd-state} for state). Also, notice that we get some NA values in education. This is actually the education level of under-school-age respondents. We omit respondents less than 18 years old in the MRP models as the (CCES) targets an adult population. Accordingly, the NAs in education response will be eventually omitted as well. The detailed code of the data pre-processing is available in the supplementary materials of this report.

\begin{table}
\caption{\label{tab:acs-response-freq}The response categories of post-stratification data.}

\centering
\begin{tabular}[t]{lr}
\toprule
Sex & percentage\\
\midrule
Male & 48.9\\
Female & 51.1\\
\bottomrule
\end{tabular}
\centering
\begin{tabular}[t]{lr}
\toprule
Race and ethnicity & percentage\\
\midrule
White alone & 67.00\\
Black or African American alone & 9.89\\
Hispanic & 14.41\\
Asian alone & 5.16\\
American Indian alone & 0.80\\
\addlinespace
Native Hawaiian and Other Pacific Islander alone & 0.15\\
American Indian and Alaska Native tribes & 0.08\\
Alaska Native alone & 0.07\\
Some Other Race alone & 0.19\\
Two or More Races & 2.24\\
\bottomrule
\end{tabular}
\centering
\begin{tabular}[t]{lr}
\toprule
Age & percentage\\
\midrule
Less than 18 years & 20.73\\
18-24 & 8.73\\
25-34 & 11.92\\
35-44 & 11.62\\
45-54 & 13.45\\
\addlinespace
55-64 & 14.64\\
65-74 & 10.95\\
75-89 & 7.00\\
90 years and over & 0.95\\
\bottomrule
\end{tabular}
\centering
\begin{tabular}[t]{lr}
\toprule
Education & percentage\\
\midrule
No high school & 27.03\\
Regular high school diploma & 18.82\\
Some college & 21.25\\
Associate's degree & 6.39\\
Bachelor's degree & 14.50\\
\addlinespace
Post-graduate & 8.99\\
NA & 3.03\\
\bottomrule
\end{tabular}
\end{table}

\hypertarget{spec}{%
\section{Model Specifications}\label{spec}}

In Chapter 2, we found that the diagnostic plots shown in many articles compare MRP estimates with other estimates. One version of this compares several MRP estimates with different model specifications. To allow us to make the same comparisons in this case study, we build five different MRP models as follows.

\textbf{Baseline model}

We begin the model fitting with the baseline model. In this model, we set the binary outcome as whether the respondents vote for Trump or not in the 2016 election. Therefore, we transform the response of \texttt{voted\_pres\_16} into a binary variable called \texttt{vote} ,i.e, if the value of \texttt{voted\_pres\_16} is ``Donald Trump'', then \texttt{vote} variable coded to ``yes'', otherwise ``no''. The NA values in \texttt{voted\_pres\_16} will stay as NA in the new \texttt{vote} variable. Tthe distribution of the baseline model's outcome variable is displayed in Table \ref{tab:vote-dist}.

\begin{table}

\caption{\label{tab:vote-dist}The distribution of answer in the outcome (vote). It will be the outcome in three models, i.e., baseline model, model with education as additional covariate, and model with more categories in race. We observe a reasonably large percentage of NA.}
\centering
\begin{tabular}[t]{lr}
\toprule
Candidate voted & percentage\\
\midrule
no & 41.00\\
yes & 29.03\\
NA & 29.97\\
\bottomrule
\end{tabular}
\end{table}

The demographic predictors used are \texttt{age}, \texttt{gender}, \texttt{state}, and \texttt{race}. As seen in Table \ref{tab:covariate-tables}, \texttt{race} has 6 categories, i.e., \texttt{White}, \texttt{Black}, \texttt{Hispanic}, \texttt{Asian}, \texttt{Native\ American}, and \texttt{All\ Other}. In the baseline model, we collapsed the \texttt{Native\ American} and \texttt{All\ Other} into \texttt{Other}. Meanwhile, the levels of \texttt{age}, \texttt{gender}, \texttt{state} stay the same in the levels displayed in Table \ref{tab:covariate-tables}. The baseline model equation is:

\begin{equation}
\begin{split}
\Pr(vote_{j[i]} = 1) & = logit^{-1}\left(\beta_0 + \alpha^{age}_{m[i]} + \alpha^{gender}_{m[i]} + \alpha^{state}_{m[i]} + \alpha^{collapsed\ race}_{m[i]}\right), \\
for\ i  & = 1, ...., n, \\
\beta_0  & \sim t(3,0,2.5) \\
\alpha^k_m  & \sim N(0,\sigma_k)
\end{split}
\label{eq:baseline-model}
\end{equation}

and \(vote_{j[i]}\) is the binary outcome (1 = yes, 0 = no) for individual \(i\) in poststratification cell \(j\). \(\beta_0\) is the intercept. \(\alpha^{age}_{m[i]}\), \(\alpha^{gender}_{m[i]}\), \(\alpha^{state}_{m[i]}\), and \(\alpha^{collapsed\ race}_{m[i]}\) are the random effects for \texttt{age}, \texttt{gender}, \texttt{state}, and \texttt{collapsed\ race}, respectively. The subscript in each coefficient represents the category of the \(i-th\) respondent, such as, \(\alpha^{collapsed\ race}_{m[i]}\) takes value from \{\(\alpha^{collapsed\ race}_{White}\), \(\alpha^{collapsed\ race}_{Black}\), \(\alpha^{collapsed\ race}_{Hispanic}\), \(\alpha^{collapsed\ race}_{Asian}\), and \(\alpha^{collapsed\ race}_{Other}\)\}. Each random effect has an independent prior distribution, such as, \(\alpha^{collapsed\ race}_{m}\) \textasciitilde{} \(N(0, \sigma^2_{collapsed\ race})\) and \(\beta_0\) \textasciitilde{} \(t(3, 0, 2.5)\). Here, we use the default prior because we only want to compare models for visualisation purpose instead of looking for the best model for estimation.

\newpage

\textbf{Model with \texttt{education} as additional covariate}

Next, we create a bigger model by adding \texttt{education} as additional covariate to the baseline model. The levels of \texttt{education} is also displayed in Table \ref{tab:covariate-tables}. Hence, the model specification is:

\begin{equation}
\begin{split}
\Pr(vote_{j[i]} = 1) &= logit^{-1}\left(\beta_0 + \alpha^{age}_{m[i]} + \alpha^{gender}_{m[i]} + \alpha^{state}_{m[i]} + \alpha^{collapsed\ race}_{m[i]} + \alpha^{education}_{m[i]}\right), \\
for\ i &= 1, ...., n.
\end{split}
\label{eq:model2}
\end{equation}

\vspace{\baselineskip}

\textbf{Model with original race categories}

This model is essentially the same with baseline model, except that there are more race categories, which are \texttt{White}, \texttt{Black}, \texttt{Hispanic}, \texttt{Asian}, \texttt{Native\ American}, and \texttt{All\ Other}. The model equation is:

\begin{equation}
\begin{split}
\Pr(vote_{j[i]} = 1) = logit^{-1}\left(\beta_0 + \alpha^{age}_{m[i]} + \alpha^{gender}_{m[i]} + \alpha^{state}_{m[i]} + \alpha^{original\ race}_{m[i]}\right), for\ i = 1, ...., n.
\end{split}
\label{eq:model3}
\end{equation}

\textbf{Model with different outcomes}

\textbf{Vote intention/preference}

This model mimicks the model in Equation \eqref{eq:model2}, except that we have a different outcome or response variable. The response here is whether the respondent intends to vote for Trump (\texttt{yes}) or not (\texttt{no}) (rather than whether they reported they voted for Trump). It is transformed from \texttt{intent\_pres\_16} variable in the CCES data to a new variable called \texttt{intent}. If the value of \texttt{intent\_pres\_16} is ``Donald Trump (Republican)'', then \texttt{intent} variable coded to ``yes'', otherwise ``no''. The NA values in \texttt{intent\_pres\_16} will stay as NA in the new \texttt{intent} variable. The distribution of observed ``no'', ``yes'', and NA in this variable is shown in Table \ref{tab:intent-dist}.

\begin{table}

\caption{\label{tab:intent-dist}The distribution of answer in the outcome (intent).}
\centering
\begin{tabular}[t]{lr}
\toprule
Candidate will be voted & percentage\\
\midrule
no & 67.76\\
yes & 29.76\\
NA & 2.47\\
\bottomrule
\end{tabular}
\end{table}

The model is specified as follows:

\begin{equation}
\begin{split}
\Pr(intent_{j[i]} = 1) &= logit^{-1}\left(\beta_0 + \alpha^{age}_{m[i]} + \alpha^{gender}_{m[i]} + \alpha^{state}_{m[i]} + \alpha^{collapsed\ race}_{m[i]} + \alpha^{education}_{m[i]}\right), \\
for\ i &= 1, ...., n.
\end{split}
\label{eq:model4a}
\end{equation}

\textbf{Party identity}

Beside vote intention, another outcome is the party identity in terms of whether the respondents identify themselves as Republican or not. This variable is derived from \texttt{pid3\_leaner} variable and referred as \texttt{party}. If the value of \texttt{pid3\_leaner} is ``Republican (Including Leaners)'', then \texttt{party} variable coded to ``Republican'', otherwise ``not Republican''. The NA values in \texttt{pid3\_leaner} will stay as NA in the new \texttt{party} variable. The distribution of this outcome variable is displayed in Table \ref{tab:party-dist}.

\begin{table}

\caption{\label{tab:party-dist}The distribution of answer in the outcome (party).}
\centering
\begin{tabular}[t]{lr}
\toprule
Party identity & percentage\\
\midrule
not Republican & 67.65\\
Republican & 32.27\\
NA & 0.08\\
\bottomrule
\end{tabular}
\end{table}

The specification of covariates is also the same with model in Equation \eqref{eq:model2}.

\begin{equation}
\begin{split}
\Pr(party_{j[i]} = 1) &= logit^{-1}\left(\beta_0 + \alpha^{age}_{m[i]} + \alpha^{gender}_{m[i]} + \alpha^{state}_{m[i]} + \alpha^{collapsed\ race}_{m[i]} + \alpha^{education}_{m[i]}\right), \\
for\ i &= 1, ...., n.
\end{split}
\label{eq:model4b}
\end{equation}

The estimates from the multilevel model is then used for the second stage of MRP, which is poststratification. As the explanation in Section \ref{overview}, poststratification is essentially taking the weighted average of the cell-wise posterior estimates with the size of each cell in the population table as the weight \autocite{GaoYuxiang2021IMRa}. For example, the poststratification estimates of people who completed High School or less and voted for Trump in the 2016 presidential election is:

\begin{equation}
\begin{split}
\theta_S = \frac{\sum_{j\in S}N_j\theta_j}{\sum_{j\in S}N_j},
\end{split}
\label{eq:poststrat-observed}
\end{equation}

where \(\theta_S\) corresponds to the proportion of 45 to 64 years old of Black Men attained High School or less (HS or Less) in Alabama who respond to ``yes'' in the \texttt{vote} variable and \(N_j\) and \(\theta_j\) are the size of cell corresponds to this sub-population category in the poststratification table and the posterior estimates of this sub-population category, respectively.

\hypertarget{prep}{%
\section{Model Preparation and Fitting}\label{prep}}

The MRP models require synchronous measurements between survey and population data. To achieve this, we need to map the survey data to the population data. In this study, the model preparation and survey-population data mapping is conducted with an R package, \texttt{mrpkit} \autocite{mrpkit}. This package allows the transparent and reproducible workflow to build MRP model, from the data mapping until the prediction stage, including the model specification setting. This package is not the product of this study but I am one of its authors. The detailed code to build the MRP models is available in the supplementary materials of this report.

After mapping the survey and population data, we can obtain a poststratification table, the first five rows of which is displayed in Table \ref{tab:post-strat-table}

\begin{table}

\caption{\label{tab:post-strat-table}First five rows of the post-stratification table}
\centering
\begin{tabular}[t]{llllllr}
\toprule
age & state & gender & collapsed\_re & original\_re & education & N\_j\\
\midrule
18 to 24 years & Alabama & Male & White & White & HS or Less & 63982.00\\
18 to 24 years & Alabama & Male & White & White & Some College & 67957.00\\
18 to 24 years & Alabama & Male & White & White & 4-Year & 8851.67\\
18 to 24 years & Alabama & Male & White & White & Post-Grad & 320.33\\
18 to 24 years & Alabama & Male & Black & Black & HS or Less & 40443.33\\
\bottomrule
\end{tabular}
\end{table}

Next, we implement a Bayesian multilevel model using \texttt{brms} \autocite{brms} to fit the model and obtain the posterior distributions of the parameters. \texttt{brms} itself incorporates either rstan or cmdstanr \autocite{cmdstanr} as the backend, which in turn wrap the probabilistic programming language Stan \autocite{stan}. We use 4000 samples of posterior distribution generated with 4 independent chains. Since this task is computationally heavy and time-consuming, we conduct it using \href{https://docs.monarch.erc.monash.edu/MonARCH/aboutMonArch.html}{Monash's High Performance Cluster (HPC)}.

\hypertarget{results-and-discussion}{%
\section{Results and Discussion}\label{results-and-discussion}}

The MRP estimates from these models will be visualised in this subsection. We will illustrate the implications of current visualisation practices and discuss the possibility for improvement using these estimates. We will divide the discussion with regards to communication and diagnostic plot as we did in in the systematic literature review (Subsection \ref{com-prac}).

\hypertarget{visualisations-for-communication-purposes}{%
\subsection{Visualisations for communication purposes}\label{visualisations-for-communication-purposes}}

One of the most widely used graphs to communicate MRP estimates is a choropleth (see Figure \ref{fig:common-plots}). Choropleth is colored, shaded, or graded to display a spatial pattern of a certain variable. For example, blue and red are used to represent states with more Democrat and Republican voters, respectively, as seen in \textcite{GhitzaYair2013DIwM}. A color gradient is also used to convey a more detailed message, for example, the state-wise MRP estimates of pro-environment opinion as seen in \textcite{EunKimSung2018Epoi}. The greener the shade, the more proportion of people support pro-environmental policy. These two examples also show the use of color with respect to the meaning that people generally perceive, i.e., green is often associated with the environment, and blue is often associated with Democrats.

\begin{figure}
\centering
\includegraphics{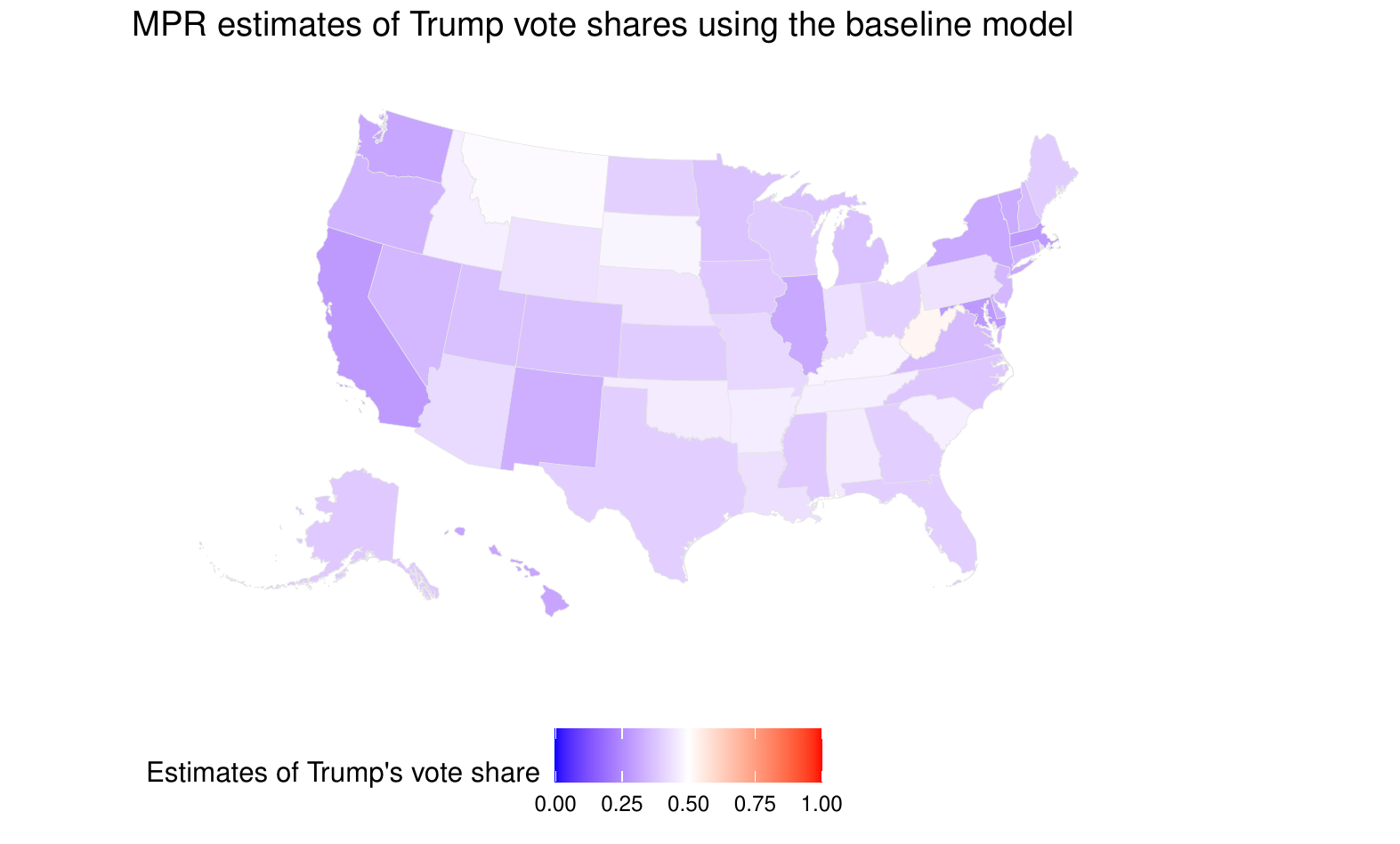}
\caption{\label{fig:choro}MRP estimates of probability of state vote for Trump in the U.S. 2016 presidential election using the baseline model. The deeper the blue shade the lower Trump's vote share in the corresponding state, while the deeper the red, the higher Trump's vote share. It is shown that the baseline model predicts that Trump has less than 50 percent vote share in almost every state in the U.S.}
\end{figure}

In this case study, we create a choropleth of MRP estimates of the probability of voting for Trump (\ref{fig:choro}) in the U.S. 2016 election using the baseline model. We create the same choropleth that is commonly shown based on our findings in the literature review.

The choropleth as seen in Figure \ref{fig:choro} conveys that the baseline model predicts that Trump has less than 50 percent of vote share in almost every state in the U.S. Regardless of whether this model has a good fit or not, the message that this graph tries to convey using color is quite easy to perceive. We can see a blue-shaded U.S. map, meaning that the Democrat candidate wins the majority of votes in most states. However, this takeaway is general, while the purpose of MRP, is to give more detailed information about sub-populations. From the map, we can see that there is only one state that has a red tint. However, the readers, especially those unfamiliar with the U.S. map, will probably not know which state this is unless the states are labeled with their name.

Another critic on the choropleth is also stated by \textcite{statgraph}. He argues that choropleth is problematic as polygons with small areas are difficult to observe. In fact, these areas sometimes carry particular information. For example, small geopolitical areas can represent a high density of population. He argues that one alternative to overcome this problem is to replace a choropleth with a cartogram in which the area is distorted so that its proportional to the value of the variable it represents. Unfortunately, there is no single visualisation among the articles reviewed that utilize this kind of visualisation.

Choropleth maps also only display point estimates, which is only one component of our analysis. Uncertainty should also be considered when visualising data, particularly estimation results, as there is always variability in these \autocite{tukey,MIDWAY2020100141,HullmanJessica2019IPoE}. In this case, a dot plot with a confidence or credible interval could be used to visualise MRP estimates, for example, as seen in \textcite{EnnsPeterK2013POit}. We can see that there is a reasonably high percentage of the usage of dot plot with uncertainty in the articles we reviewed. From the 34 dot plots found, 26 (about 76\%) of them display uncertainty. However, compared to the overall number of communication plots, the portion of the dot plot with uncertainty is only about 20\%.

\hypertarget{vis-purp}{%
\subsection{Visualisation for diagnostic purposes}\label{vis-purp}}

\textbf{Displaying Comparison of Estimation Methods}

According to \textcite{tukey}, one of the graphic's purposes is for comparison. In MRP visualisation practice, the estimates from various estimation methods are often compared. Here, we compare state-level MRP estimates with raw and weighted estimates compared to their closeness to the ground truth (actual Trump vote shares), which we obtained from R-package \texttt{ddi} \autocite{ddi}. The common aesthetic used to display this kind of purpose in the reviewed articles is a scatter plot (around 31\% of the total diagnostic plots).

\begin{figure}
\centering
\includegraphics{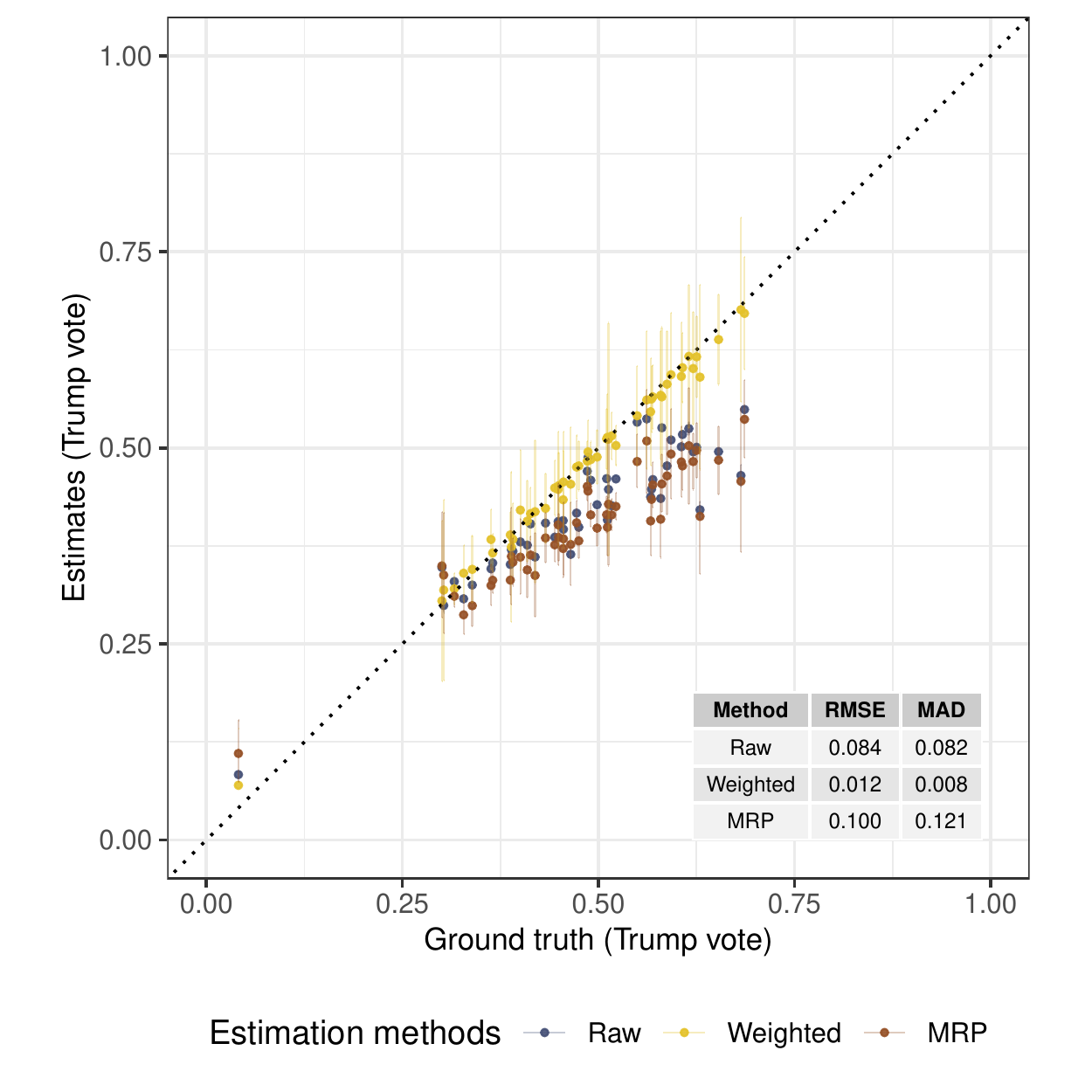}
\caption{\label{fig:scatter-est-method}Comparison between various estimates (raw, weighted, and MRP) with the actual vote share for Trump as observed in the election. The MRP model used here is the model with education as additional predictor. The points represent states with the 95 percent credible or confidence interval (depending on the method), while color represents the estimation method used. Weighted estimates is accurately predict the actual value of Trump's vote share.}
\end{figure}

There is an unwritten ``rule of thumb'' that when displaying two variables in a scatter plot, the horizontal axis displays the predictor, while the outcome is put in the vertical axis \autocite{gelmanunwin}. Regarding MRP visualisation, this ``convention'' could be translated by putting the estimates in the y-axis and the actual value in the x-axis, although the practice is sometimes interchangeable (see Figure \ref{fig:common-axis}). We also observe that some of the reviewed scatter plots show performance metrics, such as RMSE and MAE in \textcite{MengXiao-Li2018Spap}. Hence, in Figure \ref{fig:scatter-est-method}, we also display these. Most scatter plots we reviewed did not display uncertainty (see Figure \ref{fig:common-plots}). Here, we add uncertainty to each point estimates. In addition, we use color-blind-friendly color schemes to distinguish the estimation methods as mentioned by \textcite{vanderplas} and \textcite{statgraph}.

From this visualisation, we can clearly see that the weighted estimates, as seen in Figure \ref{fig:scatter-est-method}, are observed to be the most accurate. It is actually an expected result, as according to \textcite{cces_data}, the CCES's weights are poststratified to match the statewide election results.

\begin{figure}
\centering
\includegraphics{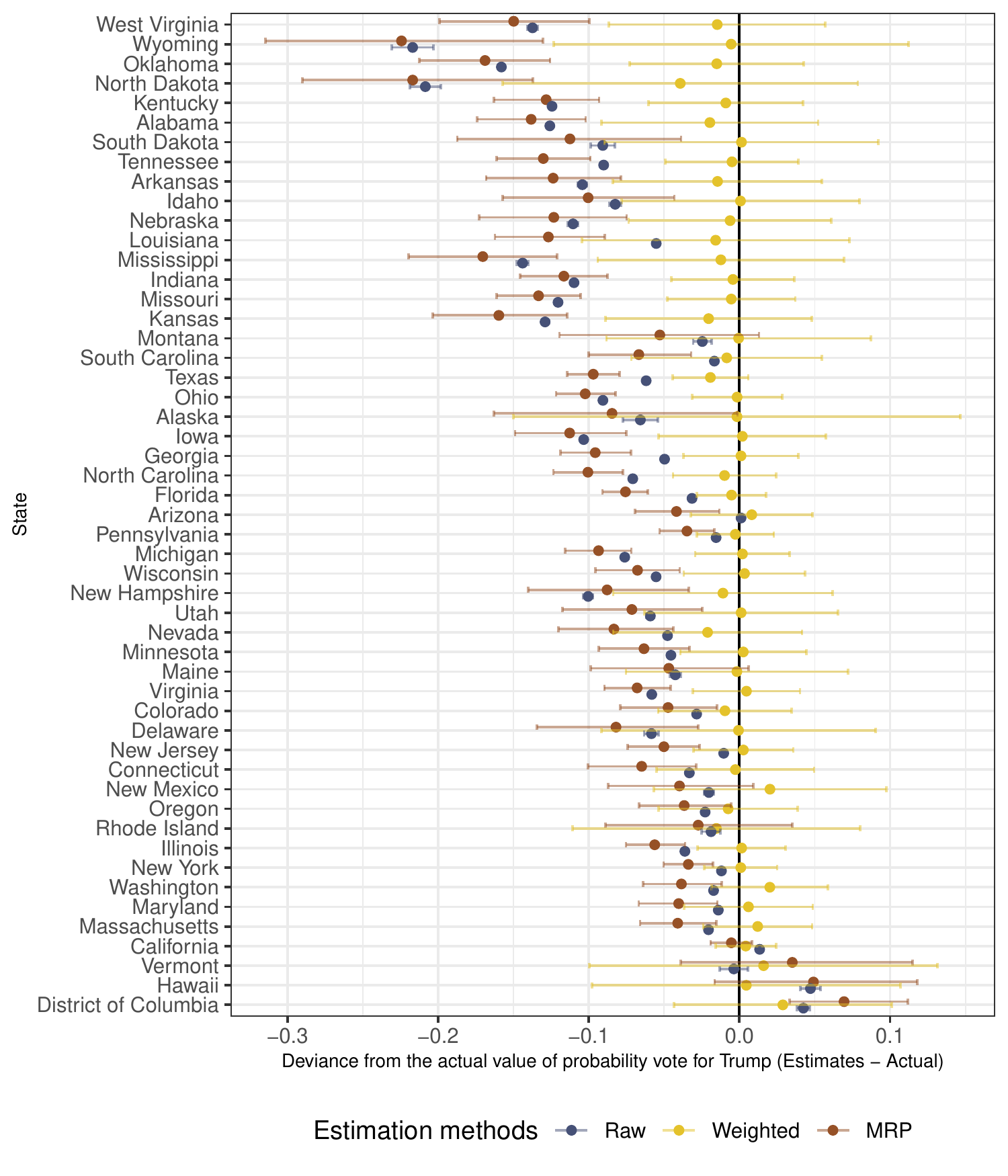}
\caption{\label{fig:dot-est-method}The deviance of estimated values from the actual value of Trump's vote share. The states in the vertical axis are ordered from states with the highest to the lowest Trump's vote share with regards to the ground truth value. The color represents the estimation methods. Again we saee that weighted estimates show the smallest deviance from the ground truth. This figure shows a pattern in which the more conservative the state, the bigger the deviance.}
\end{figure}

However, a scatter plot is appropriate when the purpose is to allow the readers to discern the general information about the relationship \emph{shape} between two variables rather than inference about \emph{individual data points} \autocite{saundra}. Hence, if the purpose is to inspect which states are least accurately estimated, the scatter plot would not be suitable. One option to help is to add labels to the points but these labels would be overlapped and hard to read in this case. Again, the dot plot could be used as an alternative to convey state-wise information, as seen in Figure \ref{fig:dot-est-method}. Here, instead of conveying the estimates, we use their deviance from the actual value of Trump's vote share, i.e., the \$Estimates - Actual~value \$. We also display the states in descending order of the actual value of Trump votes, i.e., from the most ``red'' states to the most ``blue'' states.

Using this graph, we can get the same information regarding estimate accuracy. However, we can also display other information related to the estimation error, which can then be compared across estimation methods. It shows that the more conservative the state, the higher the error. This pattern could indicate that the survey data adjustments are not sufficient to correct for sampling bias, or potentially bias between the population we poststratify to and the voting population.

\textbf{Displaying Comparison of Model Specifications}

Aside from comparing estimation methods(weighed, raw or direct estimates and mrp), we could also compare between different model specifications. Revisiting on what \textcite{WickhamHadley2015VsmR} stated, model visualisation could answer how the model fits change as the data changes. The following graphics will demonstrate this purpose.

\begin{figure}

\includegraphics{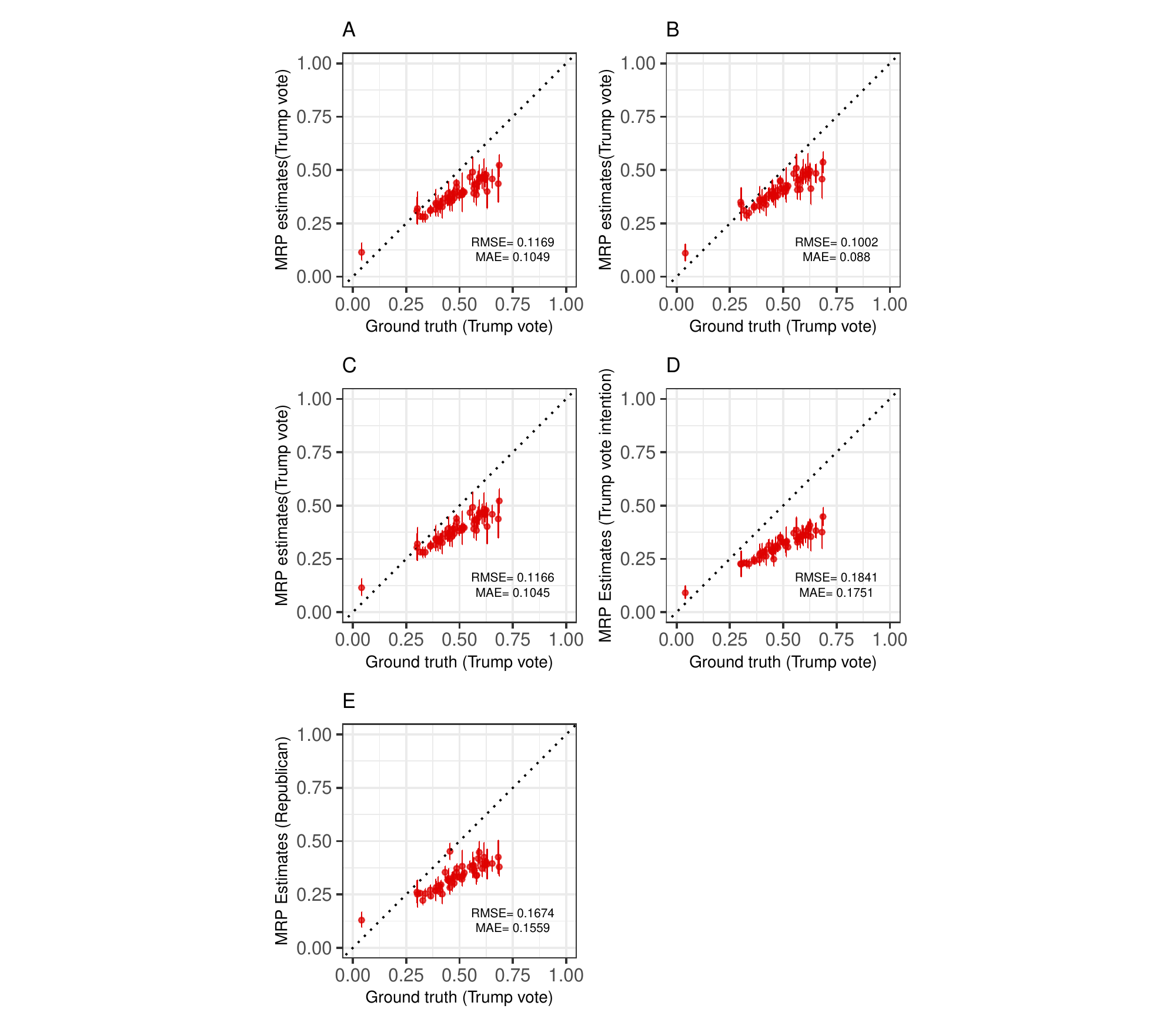} \hfill{}

\caption{Comparison between MRP estimates and the actual Trump's vote share faceted by model specification. The point represents the state. Panel A represents the fit of baseline model; B represents the model with education as additional predictor; C represents the model with more race categories; A, B, and C have the same response variable, vote, while D and E represente the model with different outcome, which are vote intention and party identity, respectively. The covariates used in model D and E are the same with the covariates of model B. We can see that all the models underestimate the actual Trump's vote share.}\label{fig:state-wise-scatter}
\end{figure}

Similar to previous plots, the comparison shown in Figure \ref{fig:state-wise-scatter} is displayed using a scatter plot. Here we compare between the MRP estimates and the actual Trump's vote share. Since there are five model specifications, we use the small-multiple principle \autocite{MIDWAY2020100141}, i.e., displaying the model fits with facets.

Using this graph, we can observe that the fit changes as the specification changes. The \(45^{\circ}\) line assists the readers in inspecting whether the fit is underestimated or overestimated. Even though almost all of the fits are underestimated, we can see that the bigger model, i.e., the model with education as an additional covariate, has a better fit than the other models (also shown by its MAE). Models in panels D and E, which are models with different outcomes, are less accurate, which is understandable as the benchmark is the actual Trump vote-share which is more aligned with the other outcome,\texttt{vote}.

In addition to estimation by small geographical area, MRP is also often used to estimate population by demographic subsets. We use violin plots to compare how the subpopulation estimates change as the model specification changes in the following visualisation. We use the violin plot as it can show the distribution of the estimates, although this plot was never observed in the articles we reviewed. It allows the reader to observe the variability and uncertainty of the rather then just the point estimates of summary statistics.

\begin{figure}
\centering
\includegraphics{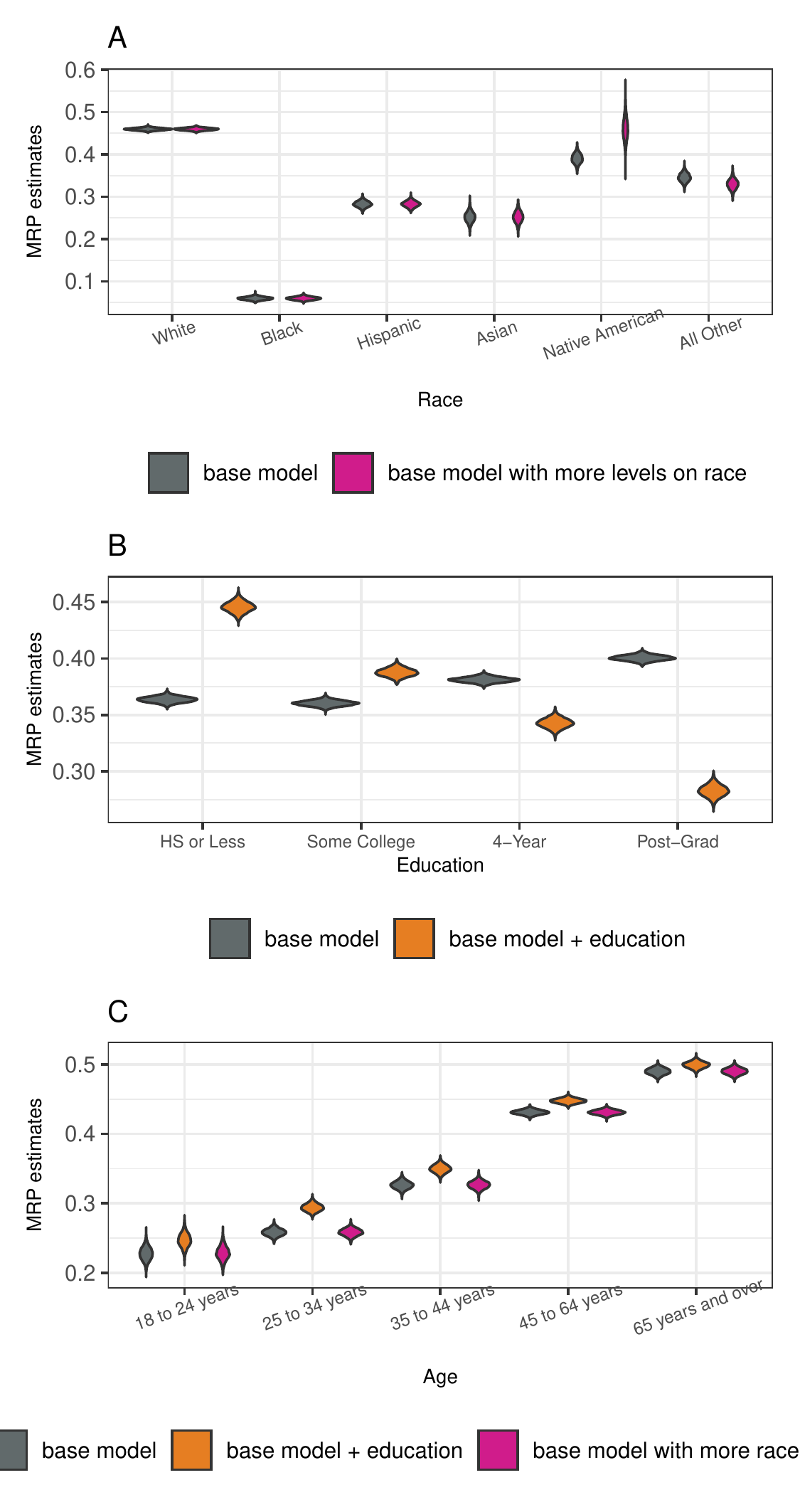}
\caption{\label{fig:violin-levels-plot}The comparison of MRP estimates by model specification. This panel shows the demographic variables estimated. Panel A, B, and C represents Trump's vote share by race categories, education level, and age group, respectively.}
\end{figure}

Figure \ref{fig:violin-levels-plot} shows the distribution of the response variable, which is probability of vote for Trump for each demographic levels regardless the geographic levels or the states where the voters live. This figure illustrates how the estimates will be different as the result of different covariates used. For example, in Panel A, the range of probability of vote for Trump of Native Americans in the model with more race categories is wider than the model that collapsed Native American and All Other as one race category. We can also see that the median of the outcome in All Other race categories is slightly different in the two models. A more pronounced difference could also be observed in Panel B which compare the baseline model with the model with education level as additional covariate. Incorporating education into the model results in a different pattern compared to the baseline model, i.e., the higher the education level, the less probability of voting for Trump. In Panel C, we can see the same trend for the three model fits, where the older age-groups tend to be more likely to vote for Trump. However, the median of model with additional adjustment variables is slightly higher in all age groups.

\textbf{Visualising Metrics}

Metrics are the performance measure of the model in estimating the ground truth. In this case, however, since the benchmark is not the actual value due to the absence of Trump's vote share in demographic levels, the term performance is not quite correct, but rather difference between these two models. We will still display graphs for these metrics, though, as an illustration of performance visualisation.

MAE and bias are predominantly used in most of the articles as the model performance criteria. Essentially, they give the same interpretation, which is how precise the model is in estimating the actual value. We also observe that correlation is frequently used in practice. Some studies also incorporate MSE/RMSE to measure their model performance.

\begin{figure}
\centering
\includegraphics{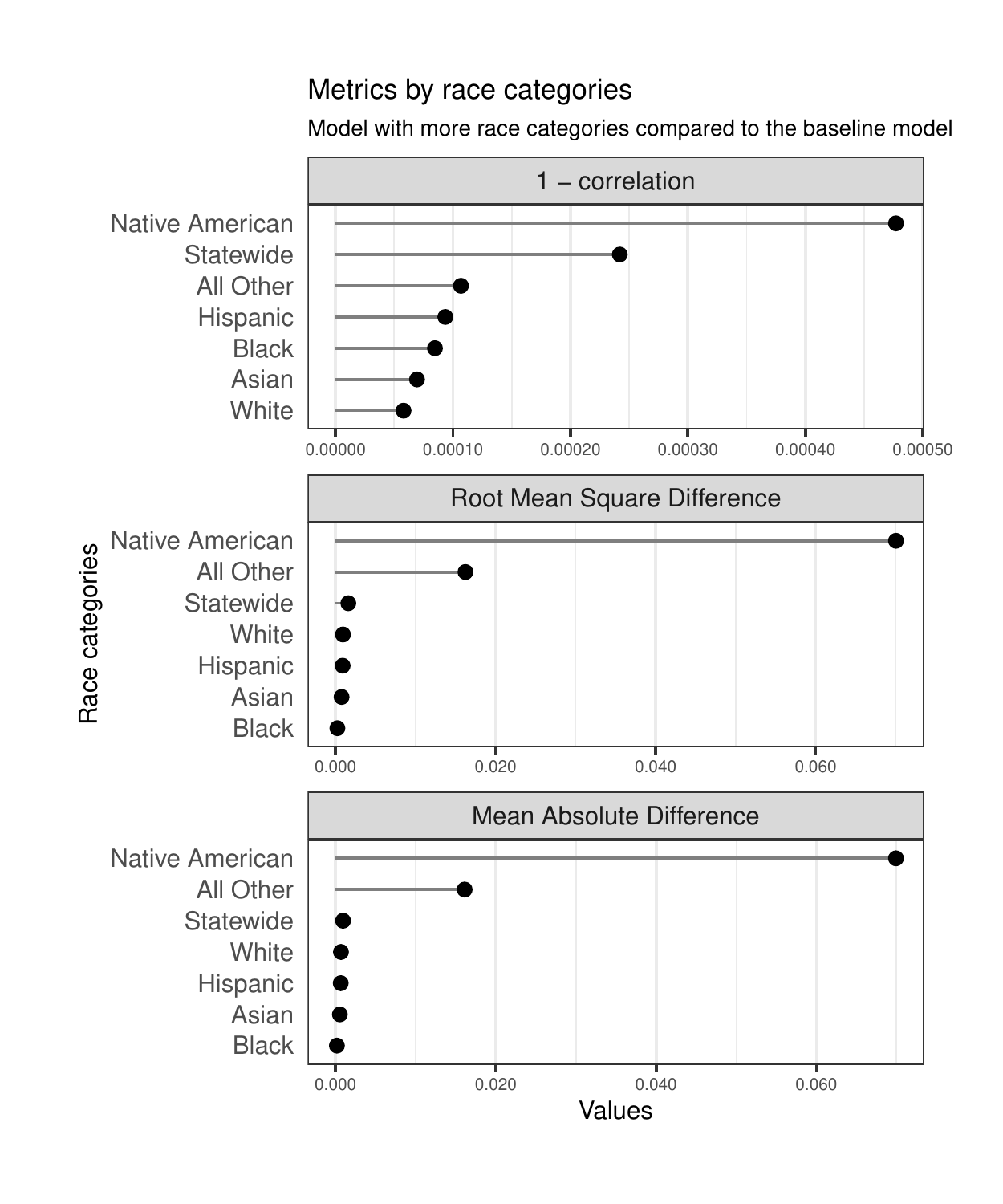}
\caption{\label{fig:cor-race-plot}Metrics of model with more race categories. This figure in only an illustration as it uses the benchmark is the baseline model, not the ground truth. Each panel shows different metrics (Correlation, Root Mean Square Deviance, and Mean Absolute Difference). The statewide categories means the state-wise metrics regardless of the race categories. Native American and All Other are the population subset with the biggest difference to the baseline model.}
\end{figure}

\textcite{WarshawChristopher2012HSWM} display correlation and MAE between state wide estimates and ground truth in a single graph by faceting it. Hence, we make a like-wise plot with a slight modification in the correlation (Figure \ref{fig:cor-race-plot}). The current practices display correlation as it is. When a graph only displays a single metric, there will be no distortion of its interpretation. However, the graph would be quite hard to read if we facet MAE or MSE/RMSE and correlation because the scales are not interpreted in the same way. For MAE and MSE/RMSE, the lower the value, the better the accuracy. In contrast, a higher correlation coefficient is more desirable. To make these scales more interpretable in the following graph, we display \(1 - correlation\) instead so that the interpretation is unidirectional. We also set the free ``scale'' so that the consistency of performance of estimates by subpopulation could be examined. Setting the display this way applies the cognitive principle of best graphical practice as stated in \textcite{vanderplas}, in which data is better presented in a way that allows the reader to compare more accurately.

From Figure \ref{fig:cor-race-plot} suggests that Native American and All Other race categories are consistently estimated to have a higher MAD and lower correlation with the baseline model. It is sensible because the baseline model collapses these categories as one covariate. This figure also illustrates how model visualisation answers whether the model is uniformly good or it is only fit for specific regions, in this case, race categories \autocite{WickhamHadley2015VsmR}.

\begin{figure}
\includegraphics[width=0.9\linewidth]{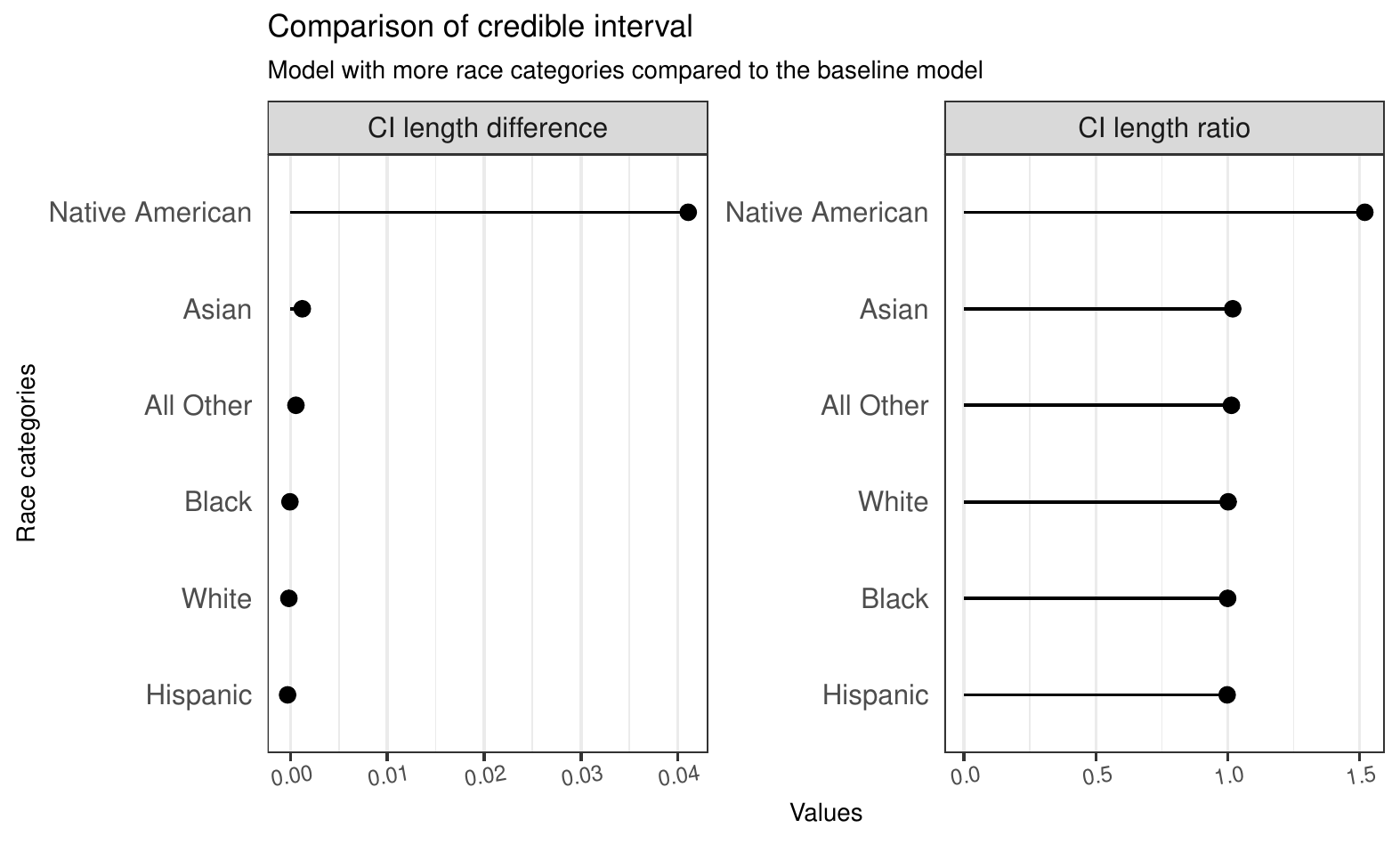} \caption{Credible interval length comparison between the model with more race categories and the baseline model. The left panel displays the mean of length difference, while the right panel display the mean of credible interval ratio. The credible interval length of Native Americans on model with more race categories is 1.5 wider than the baseline model.}\label{fig:race-ci-len}
\end{figure}

In addition to the metrics displayed in Figure \ref{fig:cor-race-plot}, we also propose alternative metrics that do not exist in the reviewed articles, namely the length of the error bar, in this case, is the 95\% credible interval. It is obtained by subtracting the 2.5\% quantile from the 97.5\% quantile of the estimates. The idea is that there is a bias-variance trade-off in MRP, and metrics, such as MAE, only take bias into account. Therefore, in Figure \ref{fig:race-ci-len}, we display the difference and ratio between the credible interval length of the model with more race categories and the baseline model. This measure will compare the variability of two model fits. If the value of credible interval length difference is near zero, then the variability of two model fits is pretty much the same. A ratio near to 1 could be interpreted in the same way.

Figure \ref{fig:race-ci-len} shows that the estimated interval of Trump's vote share in Native American categories is 1.5 wider compared to the baseline model, while other race categories generally have the same length of the credible interval with the baseline model. Hence, using this type of graph, we can sumarise that Native Americans' Trump's vote share estimate might be more uncertain when compared to other other race categories.

To sum, this demonstration shows that graphical display can help us to understand the model better. For example, the graphs have shown us that the difference in covariates or model specification could result in reasonably different estimates.

\hypertarget{ch:conclusion}{%
\chapter{Conclusion}\label{ch:conclusion}}

Graphical displays are essential to explore and understand data and model fits. They have been widely used to communicate and diagnose MRP models. However, there have been few studies formally investigating the use of visualisation within an MRP context. Therefore, in this study, we conduct a systematic literature review to understand the current practice in MRP visualisation. In addition, we perform a case study using the Cooperative and Congressional Election Study (CCES) to demonstrate the implication of current visualisation practices and explore the alternatives and possible improvements.

We find that the choropleth map is the most frequently used visualisation to communicate MRP estimates. However, it is problematic as it often hinders the information in small geographic areas and does not consider the uncertainty of estimates. Instead, we explore alternatives to display state-wise estimates using a dot plot with an error bar.
Even though it is important to show estimate uncertainty, in our literature review, we find few plots actually displayed it. We propose some alternatives to display uncertainty, for example, using a violin plot. This study also demonstrates how graphs have aided us in understanding how methods and modeling choices affect the estimates. We also use credible interval length to illustrate the bias-variance trade-off.

Naturally, this study has some limitations. In the CCES case study, none of the MRP models perform as well as the weighted estimates. However, this allowed us to demonstrate the use of visualisation to compare models that had different types of errors (such as bias or variance). Future work could compare the use of the visualisation proposed with an accurate model.

Another limitation is this study only explores alternatives to MRP visualisation in a case study. While this was useful for this study because it helped us to see the differences between visualisations, it does not conclusively provide evidence these alternatives can communicate more effectively and enhance interpretability. To get evidence of this, future work should employ careful experimentation. One example of this would be showing different types of graphs to people and seeing which aids the most accurate interpretation.

This study has provided empirical evidence on how visualisation has been performed in MRP research. One use of these findings is as a starting point to continue to explore alternative visualisations, such as those proposed in Chapter \ref{ch:case-stud}. Another use is to provide inspiration for MRP users in their own work and research. Finally, this work highlights the importance of including uncertainty which was not being included in practice.

\appendix

\hypertarget{appendix}{%
\chapter{Appendix}\label{appendix}}

\hypertarget{supplementary-material}{%
\section{Supplementary Material}\label{supplementary-material}}

All of the codes used to conduct the analysis and produce the report is available in \href{https://github.com/Dewi-Amaliah/MRP_diagnostic_plot}{this Github repository}. Particularly, the code for data wrangling and preparation can be found \href{https://github.com/Dewi-Amaliah/MRP_diagnostic_plot/blob/main/case_study/analysis/cces_acs_wrangling.Rmd}{here}, and the code for MRP preparation and visualisation can be found \href{https://github.com/Dewi-Amaliah/MRP_diagnostic_plot/blob/main/case_study/analysis/mrp_fitting.R}{here} and \href{https://github.com/Dewi-Amaliah/MRP_diagnostic_plot/blob/main/case_study/analysis/mrp_vis.R}{here}, respectively.

\hypertarget{terms}{%
\section{Terms description}\label{terms}}

There are some terms we used in Section \{com-prac\} that the readers might find unfamiliar with, especially in Figure \ref{fig:common-plots}, Figure \ref{fig:common-axis}, and Figure \ref{fig:facet-plots}. Hence Table \ref{tab:terms-desc} displays the description of those terms.

\begin{table}

\caption{\label{tab:terms-desc}Terms description used in Systematic Literature Review result}
\centering
\begin{tabular}[t]{l>{\raggedright\arraybackslash}p{30em}}
\toprule
Term & Description\\
\midrule
dot plot & Data is displayed by point/dot, one of the axis is categorical variable.\\
scatter plot & Data is diplayed by point/dot, both x and y-axis are numeric.\\
choropleth map & Thematic map coloured by the proportion of statistical variable it represents.\\
bar plot & Data is displayed by rectangular bar, one of the axis is categorical variable.\\
histogram & Similar to bar plot but number are grouped into ranges.\\
\addlinespace
density plot & Distribution of numerical variable, the y-axis is the kernel density estimates.\\
other types & Other plot types found in the reviewed articles, but the number is too few to be categorised as one category (boxplot, heatmap, bubble plot, and logit curve)\\
case & The response variable that is estimated. It sometimes displayed as faceted plot, in which each panel represents different outcome/response variable. For example, a graph contains 2 facets, A and B are the MRP estimates for opinion regarding same-sex marriage and abortion, respectively. Hence, A and B are considered as case.\\
small area & Estimates of subpopulation, geographically or demographically, for example estimates by state, county, gender, age group, education level, and religion. In some plots, it could also be another variable associates with the MRP estimates. For example, if MRP estimates used as predictor for another response variable and there is a visualisation display their relationship, then this variable is considered as small area.\\
\bottomrule
\end{tabular}
\end{table}

\hypertarget{apd-state}{%
\section{Proportion of observations by states}\label{apd-state}}

The following plots show the percentage of observations by state in CCES and ACS, respectively.

\begin{figure}
\centering
\includegraphics{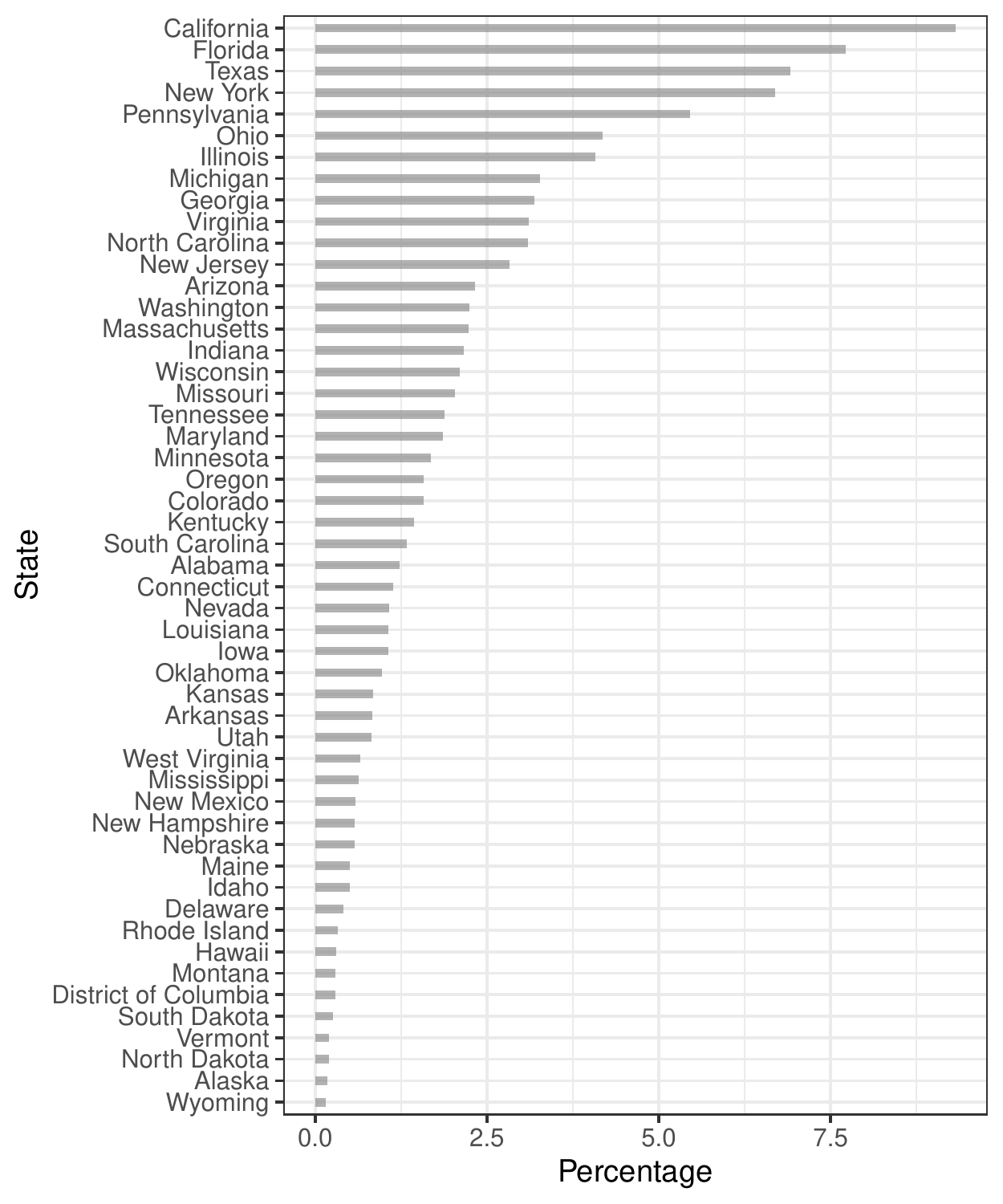}
\caption{\label{fig:state-cces}Distribution of observation in CCES data by state. The horizontal axis represents the percentage of the observations and the vertical axis represents the state ordered from the largest to lowest percentage of observations.}
\end{figure}

\begin{figure}
\centering
\includegraphics{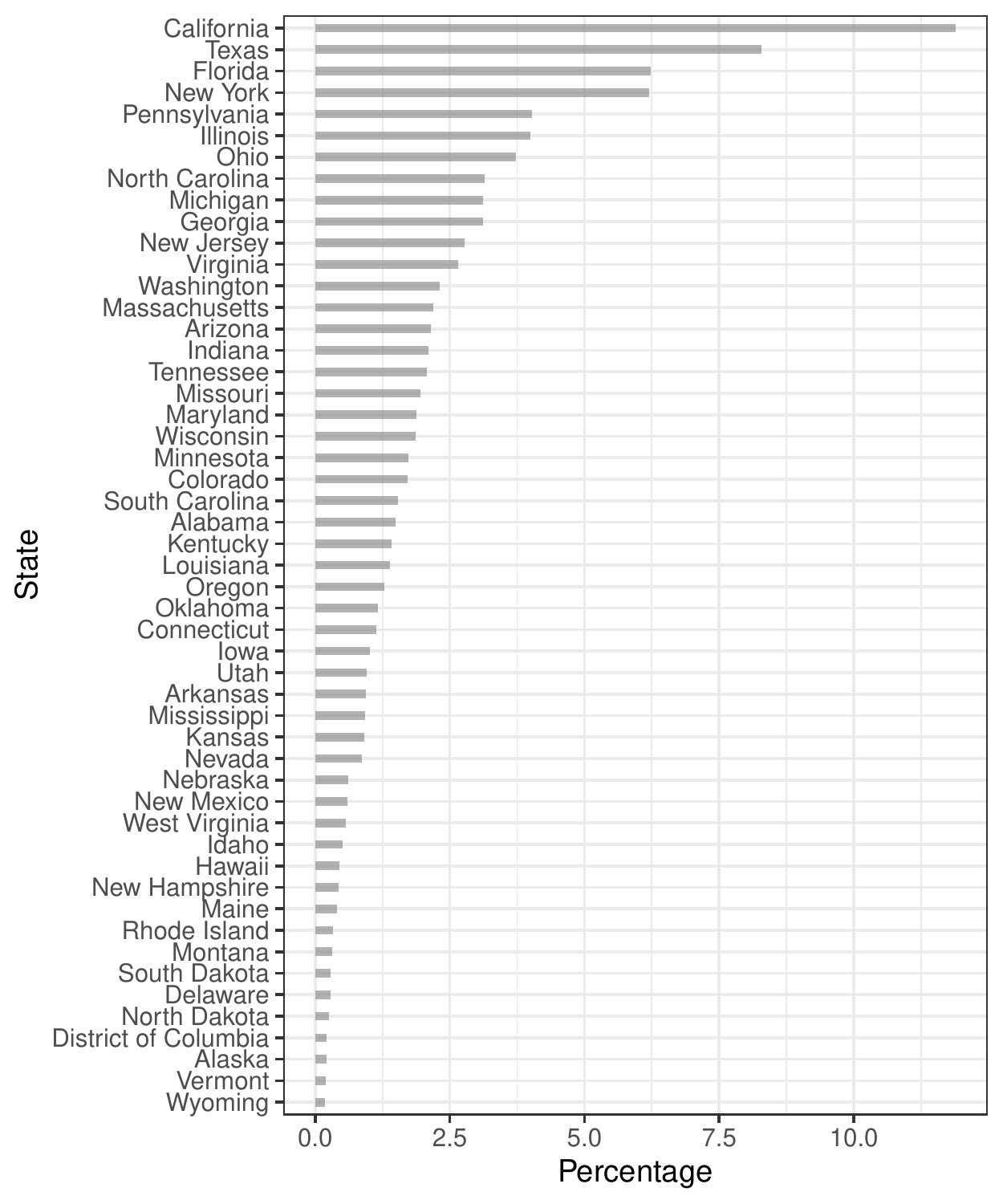}
\caption{\label{fig:state-acs}Distribution of observation in ACS data by state. The horizontal axis represents the percentage of the observations and the vertical axis represents the state ordered from the largest to lowest percentage of observations.}
\end{figure}

\hypertarget{additional-graphs}{%
\section{Additional Graphs}\label{additional-graphs}}

The following plots represent metrics and 95\% credible interval visualisation as done in Section \ref{vis-purp}.

\textbf{Education}

\begin{figure}
\centering
\includegraphics{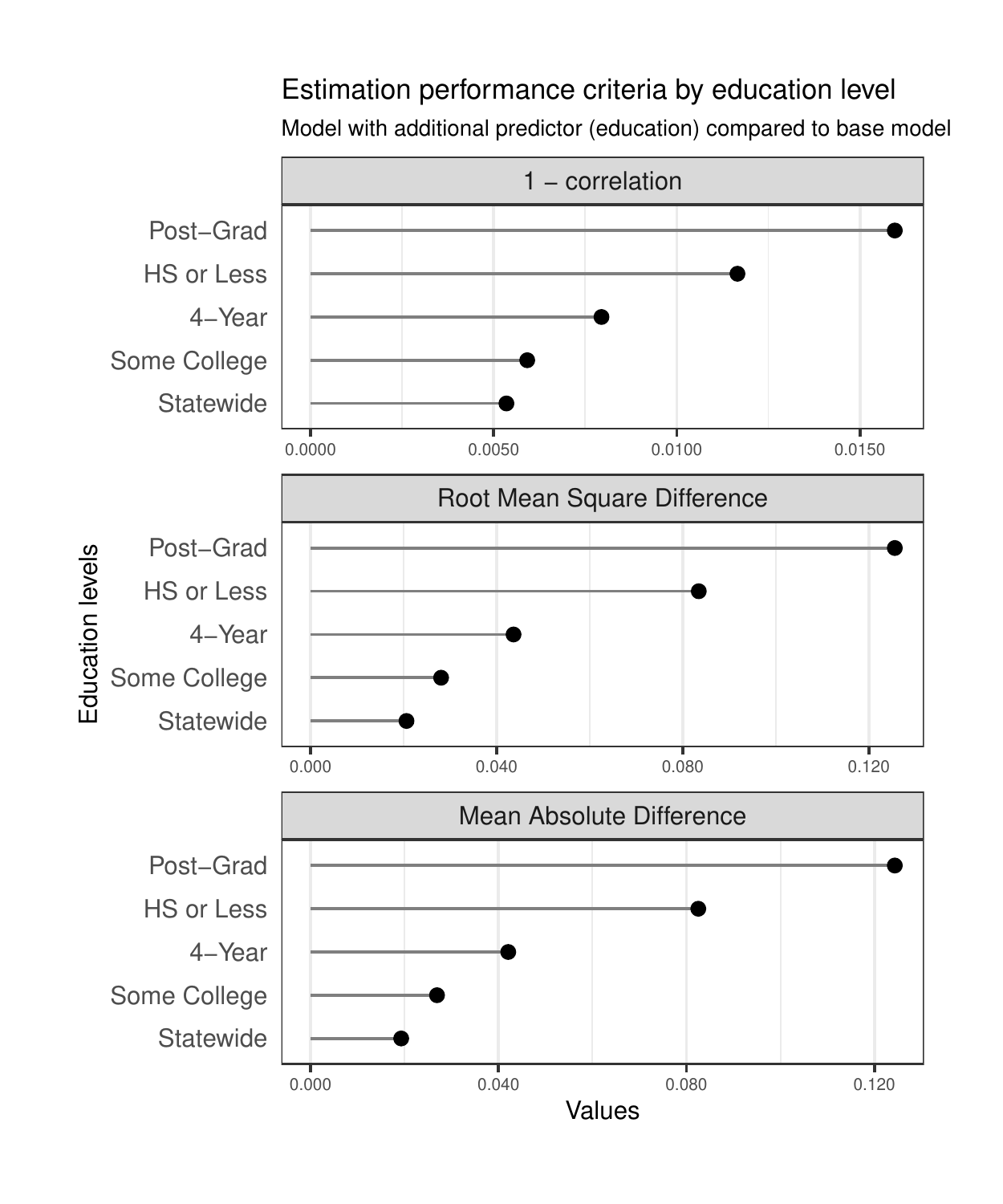}
\caption{\label{fig:apd-educ-metrics}Metrics of the model with edication as additional covariate. The benchmark is the baseline model, not the ground truth. Metrics of High school or less and Post-graduate categories are consistently have the higher deviance to the baseline model.}
\end{figure}

\begin{figure}
\includegraphics[width=0.9\linewidth]{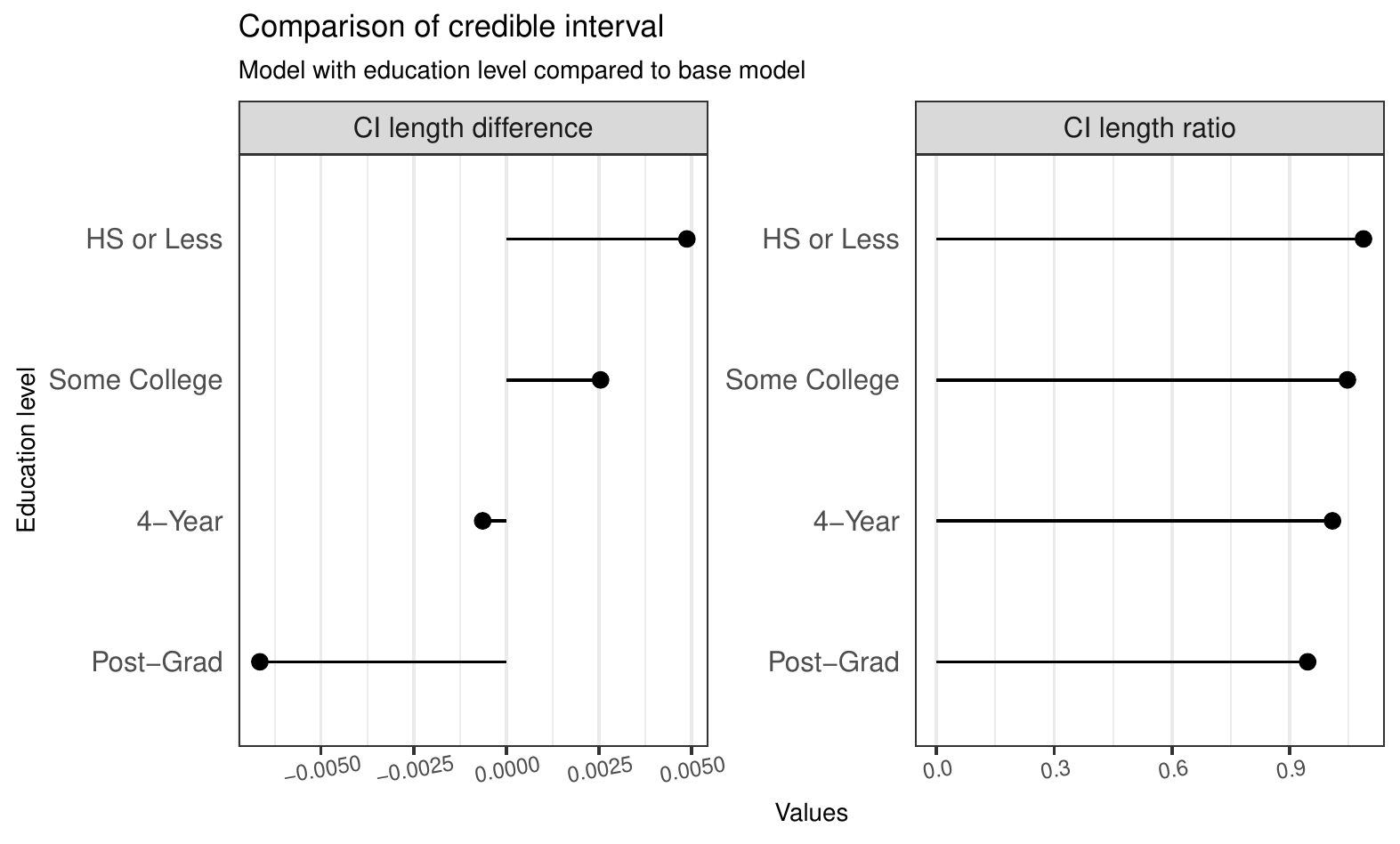} \caption{The comparison of the 95 percent credible interval length between model with education as additional covariate and the baseline model by education levels. The credible interval of bigger model for Post-graduate category is slightly narrower compared to the baseline model.}\label{fig:apd-ci-len-edu}
\end{figure}

Figure \ref{fig:apd-educ-metrics} shows the metrics and Figure \ref{fig:apd-ci-len-edu} shows the 95\% comparison of credible interval length based on education level.

\newpage

\textbf{Age (The estimation using the model with education as additional covariate)}

\begin{figure}
\centering
\includegraphics{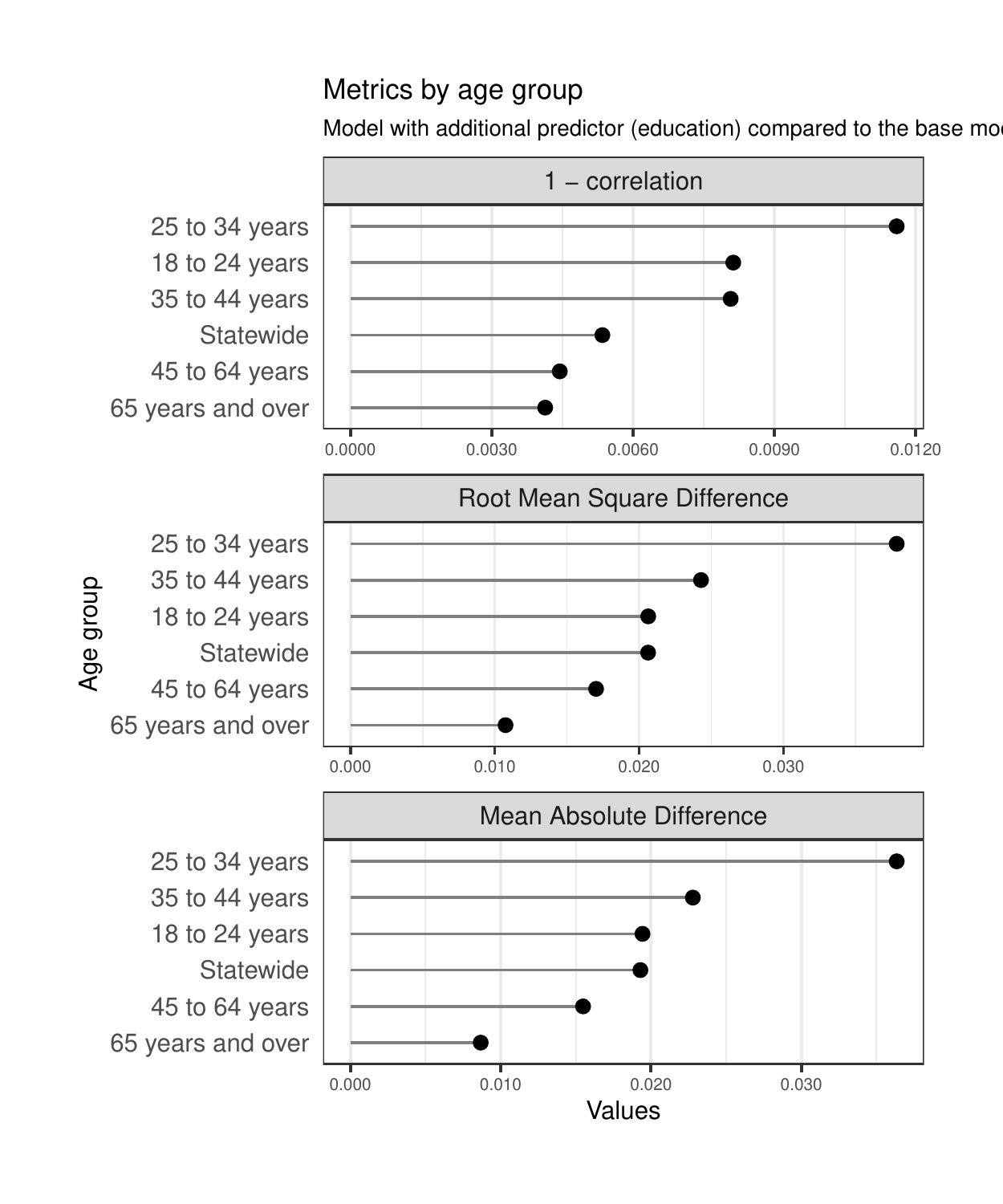}
\caption{\label{fig:apd-age-fit2}Metrics of the model with education as additional covariate. The benchmark is the baseline model, not the ground truth.}
\end{figure}

\begin{figure}
\includegraphics[width=0.9\linewidth]{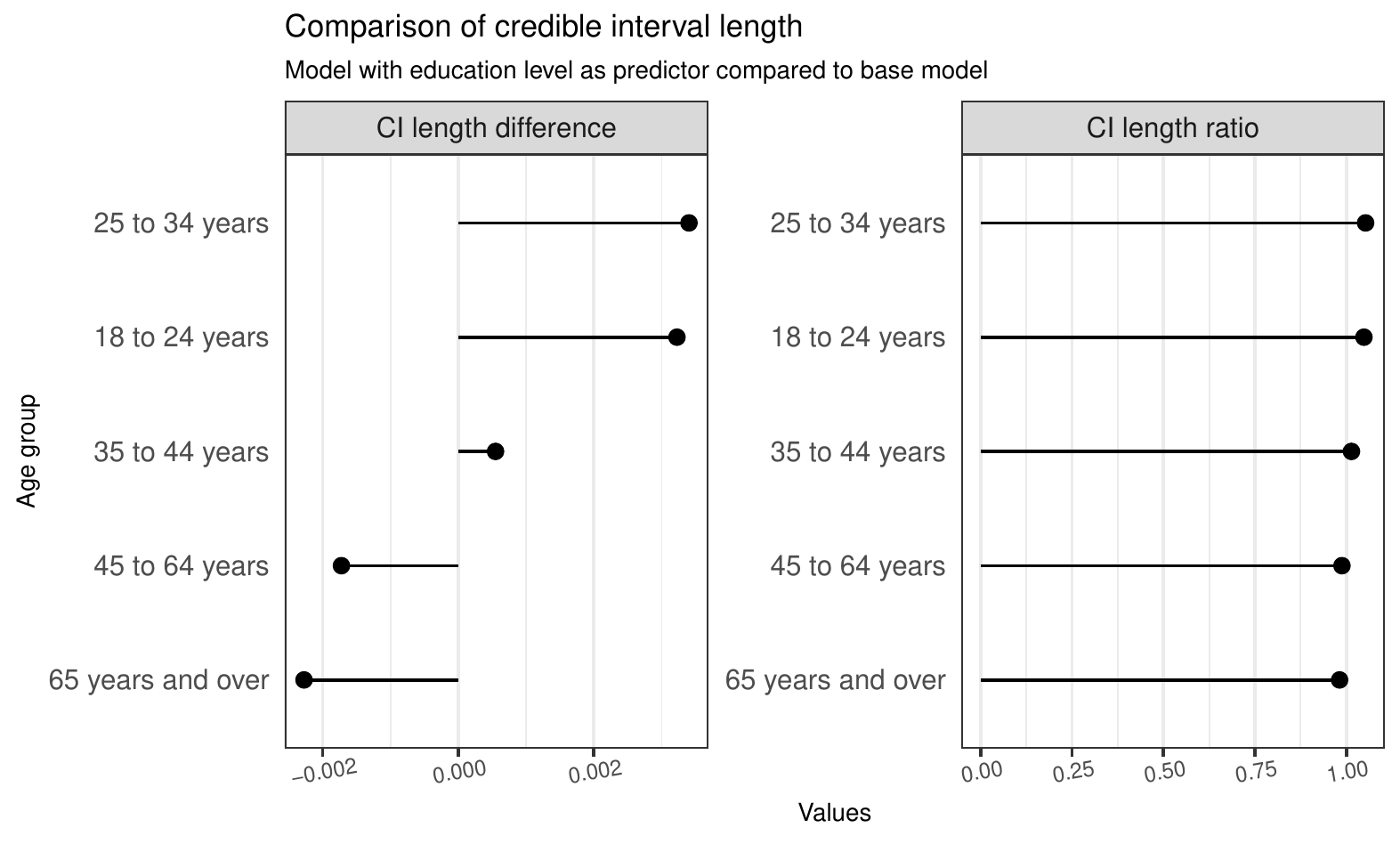} \caption{The comparison of the 95 percent credible interval length between model with education as additional covariate and the baseline model by education levels. The lenght of credible interval between the two model fits is pretty much the same.}\label{fig:apd-age-ci-fit2}
\end{figure}

Figure \ref{fig:apd-age-fit2} shows the metrics and Figure \ref{fig:apd-age-ci-fit2} shows the 95\% comparison of credible interval length based on age group (the comparison is between the model with education as additonal covariate and the baseline model).

\newpage

\textbf{Age (The estimation using the model with more race categories)}

\begin{figure}
\centering
\includegraphics{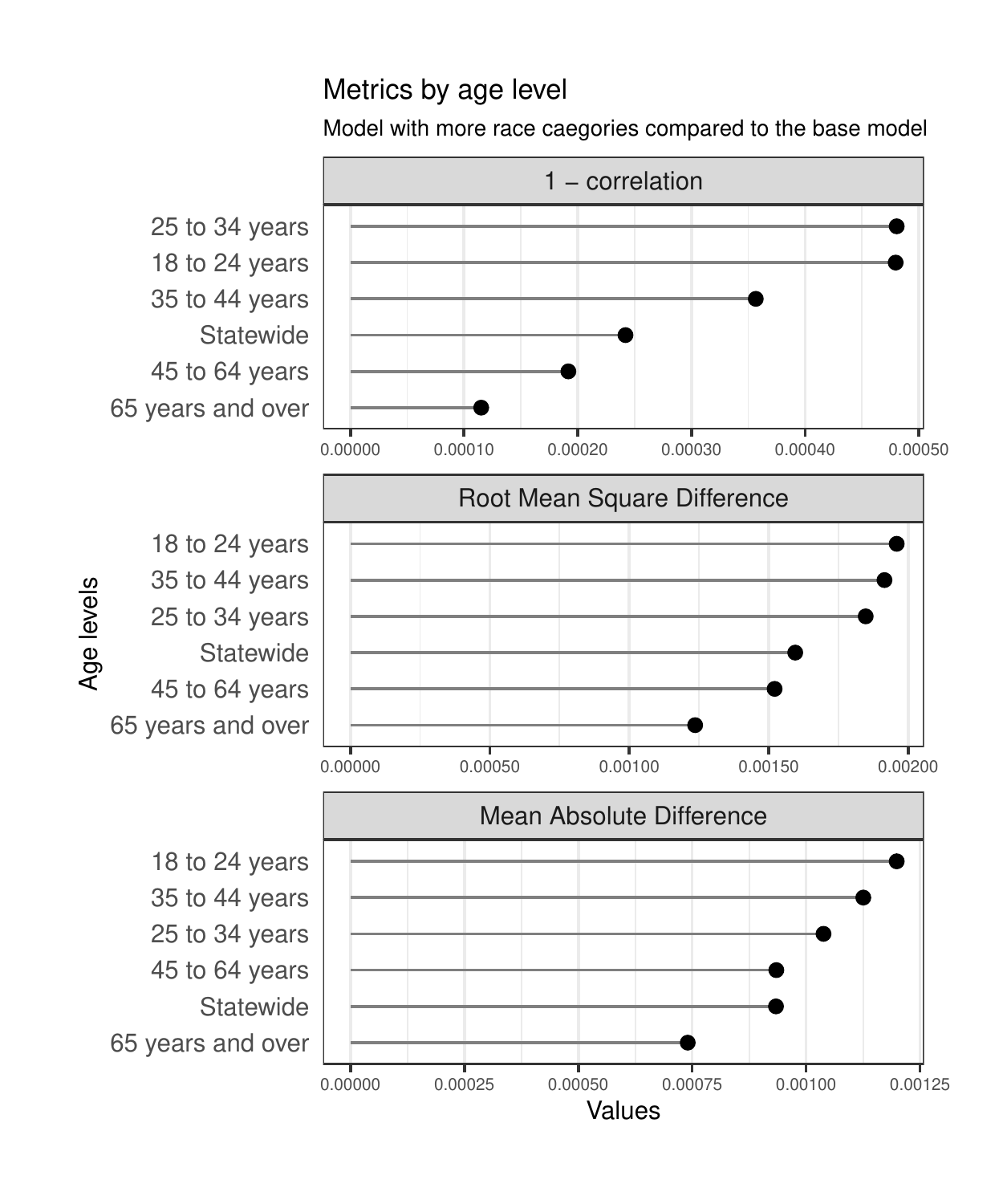}
\caption{\label{fig:apd-age-fit3}Metrics of the model with more race categories. The benchmark is the baseline model, not the ground truth.}
\end{figure}

\begin{figure}
\includegraphics[width=0.9\linewidth]{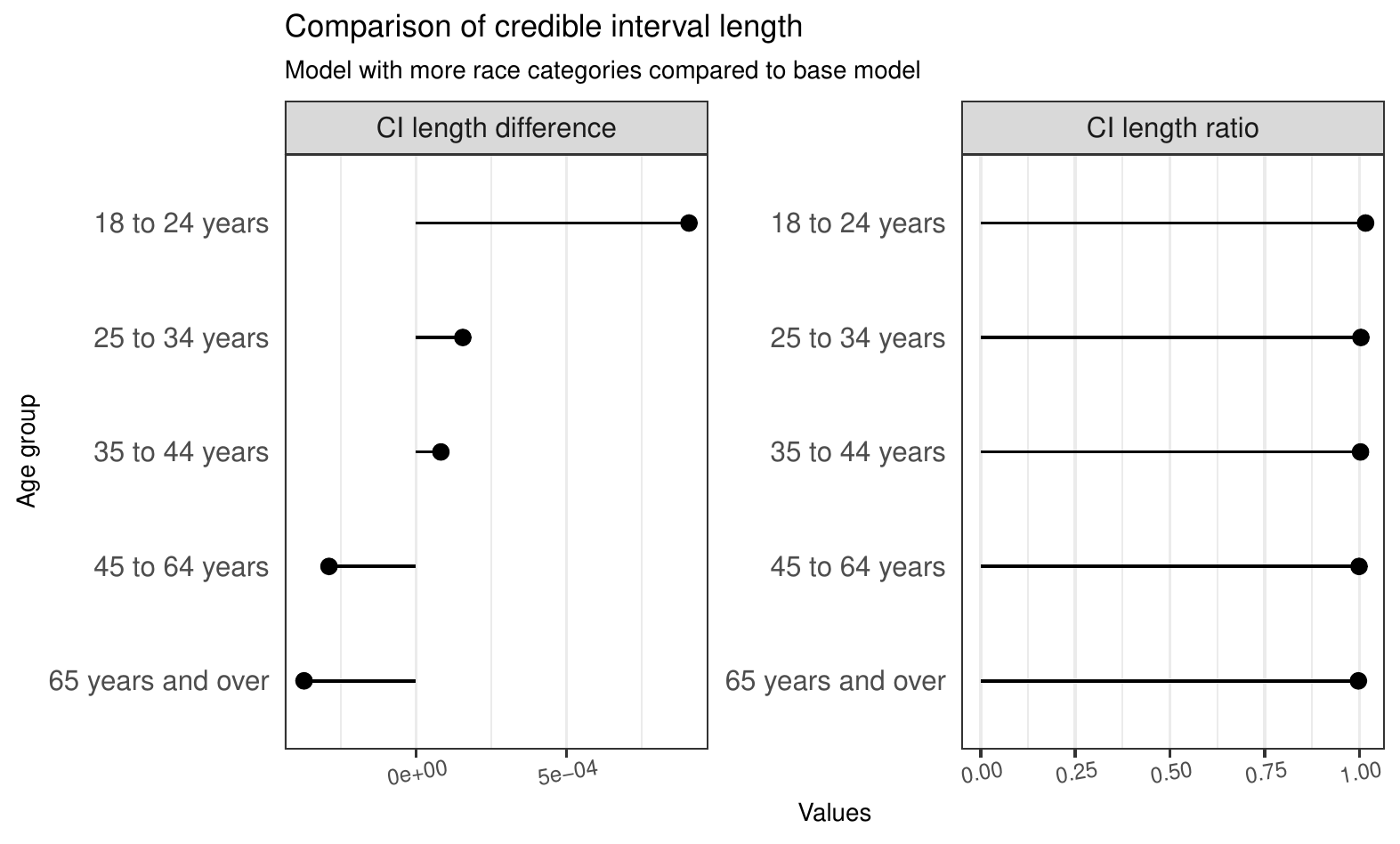} \caption{The comparison of the 95 percent credible interval length between model with more race categories and the baseline model by education levels. The lenght of credible interval between the two model fits is pretty much the same.}\label{fig:apd-age-ci-fit3}
\end{figure}

Figure \ref{fig:apd-age-fit3} shows the metrics and Figure \ref{fig:apd-age-ci-fit3} shows the 95\% comparison of credible interval length based on age group (the comparison is between the model with more race categories and the baseline model).

\printbibliography[heading=bibintoc]

\end{document}